\documentclass[12pt]{article}
\usepackage{ssi}
\usepackage{times}
\usepackage[dvips]{graphicx}

\def\PhysLett                {\mbox{Phys. Lett.}}
\def\PhysRep                 {\mbox{Phys. Rep.}}
\def\PhysRev                 {\mbox{Phys. Rev.}}
\def\PRL                     {\mbox{Phys. Rev. Lett.}}

\newcommand{\Journal}[4]     {{#1} {\bf {#2}}, {#3} ({#4})}
\newcommand{\Book}[4]        {{\it {#1}}, {#2} ({#3}, {#4})}
\newcommand{\WWWAddr}[1]     {{\tt {#1}}}
\newcommand{\hepref}[2]      {{\tt hep-{#1}/{#2}}}
\newcommand{\FermiPub}[2]    {Fermilab-Pub-{#1}/{#2}}

\newcommand{\mygraphics}[1]{%
  \includegraphics[width=\linewidth,clip=true]{#1}\vspace{-0.75cm}
}

\newcommand {\unitexp}[2] {\mbox{{#1}$^{\mathrm{#2}}$}}
\newcommand {\scinot}[2]  {\mbox{{#1}$\times$10$^{\mathrm{#2}}$}}
\newcommand {\abs}[1]     {\mbox{$\mid{#1}\mid$}}

\newcommand {\ra}         {\mbox{$\rightarrow$}}
\newcommand {\Dzero}      {\mbox{D{\O}}}

\newcommand {\Ecm}        {\mbox{$E_{CM}$}}
\newcommand {\Et}         {\mbox{$E_t$}}
\newcommand {\Etobj}[1]   {\mbox{$E_t^{#1}$}}
\newcommand {\MEt}        {\mbox{$ME_t$}}

\newcommand {\Pt}         {\mbox{$P_t$}}
\newcommand {\abseta}     {\mbox{$\mid \eta \mid$}}
\newcommand {\rphi}       {\mbox{$r-\phi$}}
\newcommand {\rz}         {\mbox{$r-z$}}
\newcommand {\dEdx}       {\mbox{d$E$/d$x$}}
\newcommand {\etaphi}     {\mbox{$\eta \times \phi$}}
\newcommand {\instL}      {\mbox{$\mathcal{L}$}}
\newcommand {\sqrts}      {\mbox{$\sqrt{s}$}}

\newcommand {\sintwoalpha} {\mbox{$\sin 2 \alpha$}}
\newcommand {\MW}          {\mbox{$M_W$}}
\newcommand {\Mt}          {\mbox{$M_t$}}
\newcommand {\MH}          {\mbox{$M_H$}}
\newcommand {\MS}          {\mbox{$M_S$}}

\newcommand{\eg}         {\mbox{\it e.g.}}
\newcommand{\etal}       {\mbox{\it et al.}}

\newcommand {\GeV}       {\mbox{GeV}}
\newcommand {\TeV}       {\mbox{TeV}}

\newcommand {\um}        {\mbox{$\mu$m}}
\newcommand {\instLunit} {\unitexp{cm}{-2}\unitexp{s}{-1}}
\newcommand {\invpb}     {\mbox{pb$^{-1}$}}
\newcommand {\invfb}     {\mbox{fb$^{-1}$}}

\newcommand {\qrk}[1]   {\mbox{${#1}$}}
\newcommand {\aqrk}[1]  {\mbox{$\bar{#1}$}}
\newcommand {\hadron}[1] {\mbox{${#1}$}}
\newcommand {\antihad}[1]{\mbox{$\bar{#1}$}}
\newcommand {\Zboson}   {\mbox{$Z$}}
\newcommand {\Wboson}   {\mbox{$W$}}
\newcommand {\Higgs}    {\mbox{$H$}}

\newcommand {\gluon}    {\mbox{$g$}}
\newcommand {\Bs}       {\mbox{$B_s$}}
\newcommand {\Jpsi}     {\mbox{$J/\psi$}}

\newcommand {\ppb}        {\mbox{$p\bar{p}$}}
\newcommand {\bbb}        {\mbox{$b\bar{b}$}}
\newcommand {\ttb}        {\mbox{$t\bar{t}$}}
\newcommand {\qqb}        {\mbox{$q\bar{q}$}}
\newcommand {\epem}       {\mbox{$e^+e^-$}}
\newcommand {\mupmum}     {\mbox{$\mu^+\mu^-$}}
\newcommand {\lplm}       {\mbox{$\ell^+\ell^-$}}
\newcommand {\vvb}        {\mbox{$\nu\bar{\nu}$}}
\newcommand {\ev}         {\mbox{$e\nu$}}
\newcommand {\muv}        {\mbox{$\mu \nu$}}
\newcommand {\tauv}       {\mbox{$\tau \nu$}}
\newcommand {\lv}         {\mbox{$\ell \nu$}}

\begin{document}
\DeclareGraphicsExtensions{.eps}

\title{  RESULTS FROM CDF AND \Dzero\ \\
	 (everything but the \hadron{B}) 
	}

\author{ Harold G. Evans\thanks{\noindent \copyright\ 2002 by Harold G. Evans}\\
	 Physics Dept., Columbia University \\
	 MailCode 5215 \\
	 538 W. 120th St., New York, NY 10027 \\[0.4cm]
	 Representing the CDF and \Dzero\ Collaborations
	}

\maketitle

\begin{abstract}%
\baselineskip 16pt 
With the start of Run II at the Fermilab Tevatron a host of new
physics opportunities are opened. In this paper we will review the
prospects for physics at the CDF and \Dzero\ experiments. Topics
ranging from QCD, to electro-weak precision measurements, to
top-quark physics, to searches for the Higgs boson and signals of
physics beyond the Standard Model will be discussed. 
B-Physics at the Tevatron is covered in a separate contribution to
these proceedings.
We will outline
how upgrades to the accelerator and the detectors make these studies
possible with precisions higher than ever achieved previously and will
show results from the first data collected in Run II.
These results
give us confidence in our ability to achieve ambitious
physics goals,
and point the way toward a bright future for the Tevatron.
\end{abstract}

\section{Introduction}
``High energy physics is a particularly exciting field right now.''
We have all heard that statement so many times that it has begun to
ring rather desperately in our ears.
In this case, however, the Bellman is right\cite{alice}.
We {\it are}
on the verge of making fundamental advances in our understanding of
questions that are of interest even to our non-physicist friends.
Why is there mass and how does it arise? Why isn't there more
antimatter? How is matter put together? 
None of these questions are addressed by the current Standard Model of
particle interactions (SM) although,
up to now,
it has succeeded in accurately predicting thousands of
experimental measurements\cite{smhewett}
(with the exception of finite neutrino masses).

Within the next decade, this situation will almost certainly
change. Our understanding of the mechanism of mass generation, which,
in the SM, is tied up with the breaking of symmetry between
electro-magnetic and weak interactions, will take a huge stride
forward with the discovery (or exclusion altogether) of the Higgs
boson, the only inhabitant of the SM zoo yet to be observed. Will this
elusive particle have all the properties predicted by the SM? Or will
its characteristics fit better with those predicted by extensions of
the SM? Perhaps it doesn't exist at all and some other mechanism will
be found to break the electro-weak symmetry? 

Answering these questions definitively is within the grasp of
experiments now running or being built. It will require a multi-prong
strategy though, with direct searches for Higgs-like particles being
complemented with predictions of the Higgs properties
using precision measurements of other electro-weak parameters (for
example, the \Wboson -boson and top-quark masses) and measurements of
rare decays (especially those of heavy particles such as $\tau$,
\qrk{b} and \qrk{t}).

Electro-weak symmetry breaking is not the only phenomenon that should
yield secrets in the coming years. The overwhelming preponderance of
matter in the observable universe is an effect that is intimately
connected with the violation of CP symmetry. All indications are that
the CP violation present in the Standard Model is not sufficient to
explain the observed asymmetry between matter and
antimatter\cite{bernreuther}. However, sources of CP violation beyond
those in the SM may well contribute here. Studies of \hadron{B}-hadron
and kaon properties as well as neutrino oscillations are crucial
to this understanding,
and again, experiments taking data now or in the near future should
clarify this question substantially.

Finally, knowing how the masses of quarks, leptons and bosons arise
and the link between this process and electro-weak symmetry breaking
still does not tell us why the proton has the mass it does. To
understand this, we must understand the intricacies of the strong
interaction. This has been a long process for which important
information will be gathered in the experiments running over the next
few years.

Looking over this list of fundamental questions it's easy to see why
the Fermilab Tevatron will be a focal point of high energy physics for
years to come. 
After a successful data taking period from 1992--1996 (Run I), which
saw, among other things,
the discovery of the top quark\cite{topdisc},
the Tevatron started a new era of data taking in March 2001
(Run II).
In Run II proton-antiproton collisions occur in the CDF and \Dzero\
detectors at a center of mass energy of 1.96 \TeV ,
which represents the
highest energy available at a collider.
We can use these interactions to study all of the questions above:
from direct Higgs searches and searches for particles beyond those
predicted in the SM, to precision measurements of SM parameters, to
studies of B-physics and CP violations, to sensitive probes of the
strong force and its theory, quantum-chromodynamics (QCD).
As you will see, these studies are expected to be consistently among
the most sensitive available with a real chance of finding something
truly groundbreaking.
To understand why, we will walk through the physics of Run II, with
the exception of B-physics, which is discussed separately in these
proceedings\cite{wuerthwein}.
We'll start with a brief discussion of the Tevatron accelerator and
the CDF and \Dzero\ detectors, especially comparing expected detector
performances. We'll then see how these detectors are used to dig out
physics signals from the large background present at a \ppb\
collider. And finally, we'll end with a brief summary of some of the
most interesting physics topics of Run II, with predictions of
expected sensitivities and indications from the first data collected
as to how the detectors are actually performing.
Unfortunately, space limitations preclude the discussion of
interesting results still coming from Run I data of the Tevatron.
So, with an eye toward the future -- let's get started.

\section{Run II at Fermilab}

\subsection{The Tevatron}
The Fermilab Tevatron accelerator facility has been substantially
modified to achieve the high luminosities required by the physics
goals of Run II, which
began officially in March of 2001. The main changes with respect to
previous Tevatron running are in the center-of-mass energy of the
\ppb\ collisions, which has been increased from 1.8 \TeV\ to 1.96
\TeV\ between Runs I and II,
and in the instantaneous luminosity,
which should increase by more than two orders of magnitude with
respect to the values achieved in Run I. Some of the factors
contributing to these changes are detailed in Table
\ref{table:tevatron}
(see Ref. \citenum{tevupgrade}).

Because of these accelerator changes physics prospects at Run II are
improved in two ways. Obviously, the increase in the amount of data
available made possible by increased Run II luminosities
will allow new analyses to be performed and will increase the statistical
precision of old ones. However, the small increase in CM energy over
Run I actually results in a substantial increase in cross-section for
several interesting physics channels. For example, cross-sections for
\Wboson /\Zboson , top quarks and jets with \Pt\ $>$ 400 GeV will
increase by factors of 1.1, 1.35 and 2, respectively.

By July, 2002 the Tevatron had delivered approximately 50 \invpb\ of
data to CDF and \Dzero . Of this data the experiments recorded
10--20 \invpb ,
which was used to produce the
results presented at the
ICHEP02 conference in Amsterdam\cite{ichep02}
upon which this paper is based.
In the future, Run II data taking is foreseen to happen in two main
stages -- Run IIa, where
approximately 2 \invfb\ of data will be collected, and
Run IIb, where a total integrated luminosity of 10--15 \invfb\ is
hoped for. The exact timing of the transition between Runs IIa and IIb
will be determined by degradation of the CDF and \Dzero\ silicon
detectors with radiation dose accumulated and is expected to occur
around 2005-2006. Details of the machine goals for Runs IIa and IIb
are given in Table \ref{table:tevatron}.

\begin{table}[h]
\begin{center}
\caption{A comparison of the Run I and Run II parameters at the
  Fermilab Tevatron.}
\label{table:tevatron}
\begin{tabular}{|l|c|cc|c|}
\hline
\hline
  & {\bf Run Ib} & \multicolumn{2}{|c|}{\bf Run IIa} & {\bf Run IIb} \\
  & typical & \ra\ July 02 & goal & goal \\
\hline
\hline
  Years & 92--96 & \multicolumn{2}{|c|}{01--05} & 06--LHC \\
\hline
  \Ecm  [\TeV ] & 1.8 & 1.96 & 1.96 & 1.96 \\
\hline
  Bunches (\hadron{p}$\times$\antihad{p})
    & 6$\times$6 & 36$\times$36 & 36$\times$36 & 36$\times$36(140$\times$103) \\
  Bunch Spacing [ns] & 3500 & 396 & 396 & 396(132) \\
\hline
  Total protons      ($\times$\unitexp{10}{12}) & 1.4 &     & 9.7 & 9.7(38) \\
  Total anti-protons ($\times$\unitexp{10}{12}) & 0.3 &     & 1.1 & 3.4(9.6) \\
\hline
  $<$Interac's/X'ing$>$
    & 2.5  & $<$1 & 2.3 & 5.5(3.7) \\
\hline
  Inst. Lumi. ($\times$\unitexp{10}{32} \instLunit ) 
    & 0.16 & 0.2  & 0.86 & 2.0(4.1) \\
  Integ. Lumi [\invfb ]
    & 0.125    & 0.01 & 2    & 15 \\
\hline
\hline
\end{tabular}
\end{center}
\end{table}

\subsection{The Detectors}
To take full advantage of the physics possibilities in Run II, both
the CDF and \Dzero\ collaborations have made major upgrades to their
detectors\cite{cdfweb,d0web}.

\begin{figure}[ht]
\begin{center}
  \mygraphics{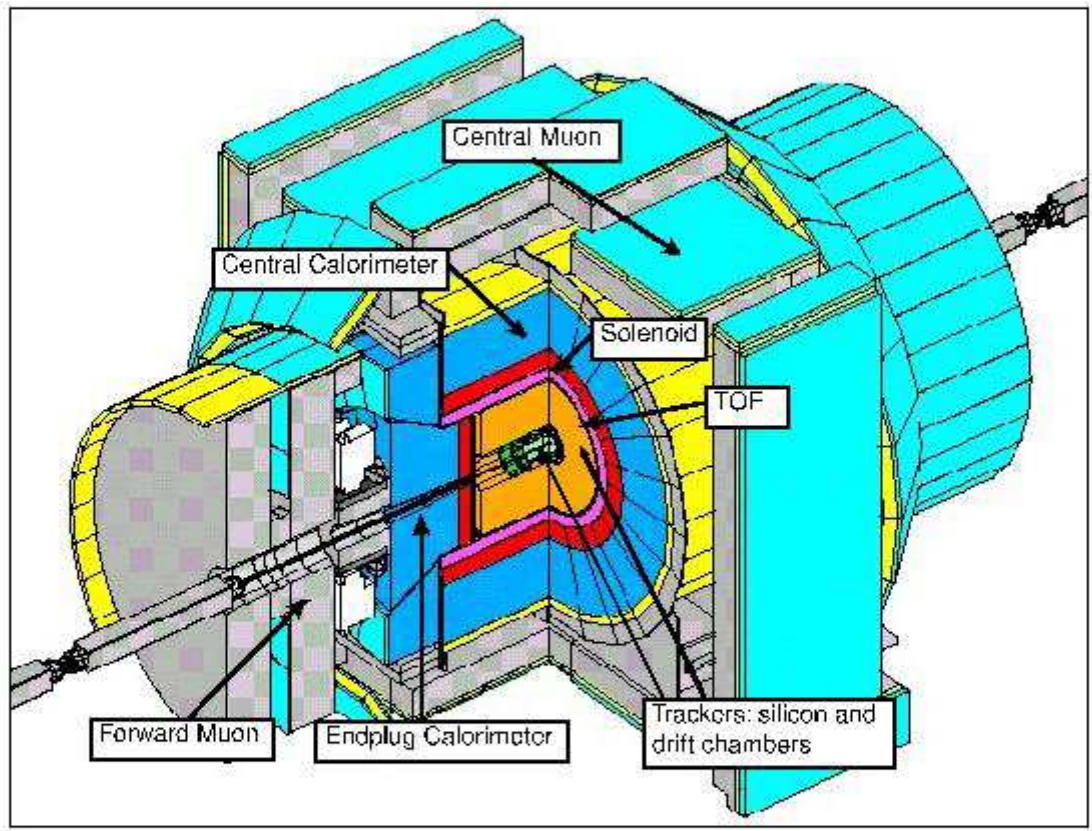}
  \caption{The CDF Run II detector.}
  \label{fig:cdfdet}
\end{center}
\end{figure}

CDF has replaced their Run I silicon detector with a new device
providing 3D tracking up to \abseta $<$2.\footnote{The pseudo-rapidity
  is defined as $\eta \equiv -\ln \tan(\theta/2)$.}
Further improvements in tracking come from a new, faster drift
chamber with 96 layers (COT) and 
new Time-of-Flight (TOF) detector.
They have also significantly enhanced their capabilities in the
forward region with a new plug calorimeter and a new forward muon
system. Finally, they have upgraded their trigger system and added a
new track trigger at Level-1, based on information from their drift
chamber as well as constructing a new impact parameter trigger at
Level-2 using data from their silicon detector. A cut-away view of the
CDF Run II detector is shown in Figure \ref{fig:cdfdet}.

\begin{figure}[ht]
\begin{center}
  \mygraphics{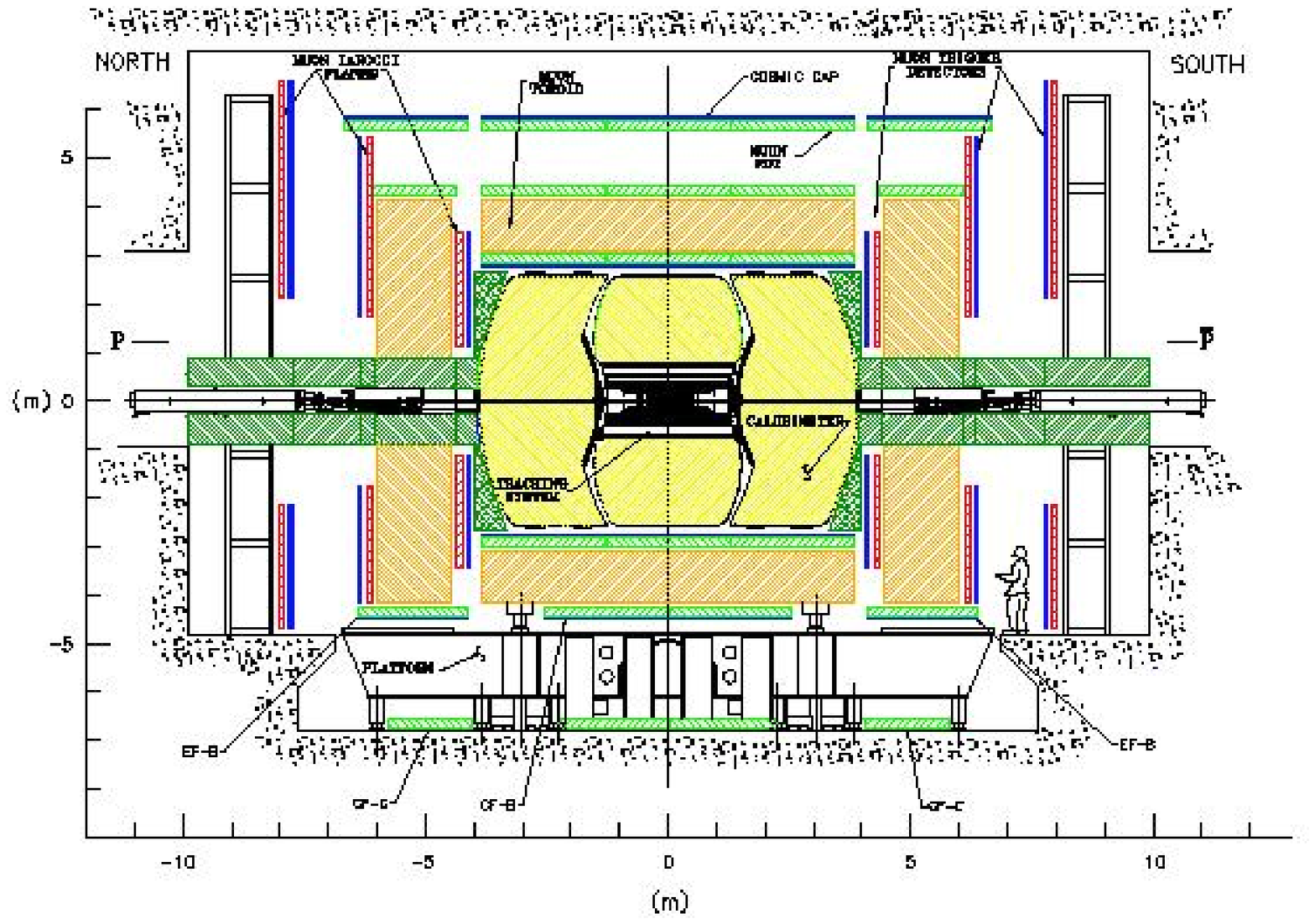}
  \caption{The \Dzero\ Run II detector.}
  \label{fig:d0det}
\end{center}
\end{figure}

The \Dzero\ upgrade was also an ambitious project. The old,
non-magnetic tracking system was completely replaced and now includes
a silicon micro-vertex detector with 3D readout and a central tracker
(CFT) using 8 super-layers of scintillating fibers immersed in a 2.0 Tesla
axial magnetic field. Because of the increased amount of material in
the tracking system, pre-shower detectors have been added in the
central and forward regions. The \Dzero\ uranium-liquid argon
calorimeter has been retained, but its readout electronics have been
completely replaced. The muon system in the central region is also
largely unchanged from Run I. However, trigger scintillator counters
have been added and the muon system in the forward region has been
completely replaced. Finally, totally new trigger and data acquisition
systems have been installed. A diagram of the \Dzero\ Run II detector
is shown in Figure \ref{fig:d0det}.

Both detectors are now operating quite well, with only some aspects of
the \Dzero\ trigger system remaining to be commissioned. Detector
performance is now approaching design goals for many of the
sub-systems. A summary of some of the performance goals in Run II is
given in Table \ref{table:detperf}. As can be seen, the detectors have
been built to the highest standards and are expected to have largely
similar capabilities. Several differences are worth mentioning,
however, as they highlight the differences in many of the analyses that will
be performed by the collaborations. It should be emphasized that these
performance differences tend to be rather small with both
experiments expected to be able to measure a large range of phenomena with
similar precision.

One of CDF's main strengths is their superb tracking system. Their
excellent momentum and vertex reconstruction resolution give them an
advantage in several areas of B-physics where final states must be
reconstructed using several relatively low momentum tracks.
\Dzero , on the other hand, has very strong calorimetry and muon
detection covering a larger solid angle than CDF's. This is especially
helpful in some analyses of physics beyond the standard model where
muon, electron and jet acceptances are important.
Finally, CDF and \Dzero\ have substantial differences in the
philosophy of their trigger systems. While both experiments use a
three-level trigger, with the first level using custom
hardware, the second using special purpose CPUs and the third using a
farm of PCs, an important difference can be found at level-1. Mainly
because of the choice of the silicon detector readout chip, CDF can
issue level-1 accepts at rates up to 50 kHz, while the \Dzero\ level-1
system is limited to approximately 5 kHz. The main consequence of this
difference is that CDF can construct a low-\Pt\ track trigger aimed at
such B-physics topics as \sintwoalpha\ and \Bs\ mixing using
fully hadronic decays. This extended level-1 rate capability is not
expected to give much advantage in such high-\Pt\ topics as vector
bosons, top and Higgs physics and physics beyond the standard model,
though, since signal rates here are already quite low.

Finally, a word about coordinate systems. Both CDF and \Dzero\ use
coordinate systems with the $z$-axis pointing along the proton beam
direction, the $x$-axis away from the center of the ring and the
$y$-axis pointing up. Azimuthal angles (in the $x-y$, or transverse
plane) are generally denoted as $\phi$, with $r$ measuring the
distance from the beam line in this transverse plane. The polar angle,
$\theta$, is measured from the $z$-axis and pseudo-rapidity, $\eta$,
is defined using it, as described above.

\begin{table}
\begin{center}
\caption{A comparison of expected \Dzero\ and CDF detector
  performances in Run II.}
\label{table:detperf}
\begin{tabular}{|lc|c|c|}
\hline
\hline
  & & {\bf \Dzero } & {\bf CDF} \\
\hline
\hline
  \multicolumn{4}{|c|}{\bf Tracking System} \\
\hline
  Technologies     &     & silicon, scintillating fibers
                         & silicon, drift chambers \\
  Magnetic Field   & [T] & 2.0 & 1.4 \\
  \abseta\ accept. &     & $<$3.0(Si), $<$1.7(CFT) 
                         & $<$2.0(Si), $<$1.0(COT) \\
  Radii & [cm]           & 2.8--10.0(Si), $<$52(CFT)
                         & 1.6--10.7(Si), $<$132(COT) \\
  $\delta$\Pt /\Pt &[\%] & 2 $\oplus$ 0.2\Pt
                         & 0.7 $\oplus$ 0.1\Pt \\
  Impact param res  & [\um ] & 13 $\oplus$ 50/\Pt
                             & 6 $\oplus$ 22/\Pt \\
  Primary vtx res   & [\um ] & 15--30(\rphi )
                             & 10--35(\rphi ) \\
  Secondary vtx res & [\um ] & 40(\rphi ), 80(\rz )
                             & 14(\rphi ), 50(\rz ) \\
  Mass res \Jpsi \ra \mupmum & [MeV] & 27 & 15 \\
  Particle ID       &        & pre-shower & \dEdx , TOF \\
\hline
  \multicolumn{4}{|c|}{\bf Calorimetry} \\
\hline
  Technologies     &            & uranium-liquid Ar
                                & lead-scint./prop.-chambers \\
  \abseta\ accept. &            & $<$4.0 & $<$3.6 \\
  Granularity      & (\etaphi ) & 0.1$\times$0.1 & 0.1$\times$0.26 \\
  EM res.          & [\%]       & 14/$\sqrt{E}$ & 16/$\sqrt{E}$ \\
  Jet res.         & [\%]       & 80/$\sqrt{E}$ & 80/$\sqrt{E}$ \\
\hline
  \multicolumn{4}{|c|}{\bf Muon System} \\
\hline
  Technologies     &            & drift tubes, scintillator
                                & drift chamb's, scintillator \\
  Magnetic Field   & [T]        & 1.8    & 0 (central) \\
  \abseta\ accept. &            & $<$2.0 & $<$1.5 \\
  $\phi$ coverage  & [\%]       & $>$90  & $>$80 \\
  Shielding        & int. len.  & 12--18 & 5.5--20 \\
  Standalone $\delta$\Pt /\Pt   &[\%] & 18 $\oplus$ 0.3$P$ & --- \\
\hline
  \multicolumn{4}{|c|}{\bf Trigger System} \\
\hline
  Hardware         & L1         & custom electronics & custom electronics \\
                   & L2         & custom CPUs        & custom CPUs \\
                   & L3         & PC farm            & PC farm \\
  Accept rate      & L1 [Hz]    & 5000   & 50000 \\
                   & L2 [Hz]    & 1000   & 300 \\
                   & L3 [Hz]    & 50     & 50 \\
\hline
\hline
\end{tabular}
\end{center}
\end{table}

\section{Physics and How to Find It}
Physics of interest in proton-antiproton interactions at the Tevatron
comes from the ``hard scattering'' of a pair of (anti)quarks or
gluons that make up the \hadron{p} and \antihad{p}. 
Because of the large strong coupling constant, this hard
scattering is governed almost exclusively by QCD, which is
well-understood at these energies,
and generally results in the production of light quarks or gluons with
subsequent gluon radiation.
These final-state partons then hadronize to produce the particles
observed in the detector.

Such QCD processes are referred to as
``low-\Pt '' because the transverse momenta of
the objects produced in the hard scatter tend to have \Pt 's small
compared to the beam energy.
More rare events, such as
vector boson, top quark and Higgs production as well as signals of physics
beyond the standard model are called ``high-\Pt '' because they
contain objects with relatively large \Pt .

Also present in any hard scattering event at the Tevatron are the
remnant partons of the proton and antiproton, which also hadronize to
form jets of particles, referred to as the ``underlying event'',
traveling basically along the beam direction.
Occasionally, some of the particles from the underlying event are
produced with relatively large transverse momenta and contaminate the
products of the hard scattering, confusing event classification.

An idea of the problems and opportunities facing those physicists
studying high \Pt\ 
physics at the Tevatron can be obtained by examining the first three
columns of Table \ref{table:tevphys}.
Event rates are high enough that we will be able to record significant
samples of some of the most interesting physics processes. However,
the QCD multi-jet cross-section is huge -- 10 orders of magnitude
higher than top production, for example.

\begin{table}
\begin{center}
\caption{Cross-sections, rates (at \instL\ = \scinot{2}{32} 
  \instLunit ), and event characteristics of various physics processes
  at the Tevatron in Run II.}
\label{table:tevphys}
\begin{tabular}{|l|cc|cccc|}
\hline
\hline
  {\bf Mode} & {\bf X-Sect} & {\bf Rate} 
    & $\mathbf{< E_t^{jet} >}$ & $\mathbf{< E_t^{lept} >}$ 
    & $\mathbf{< ME_t >}$ & {\bf Displ. V.} \\
  & & & [\GeV ] & [\GeV ] & [\GeV ] & [mm] \\
\hline
\hline
  Inelastic \ppb & 50 mb & 10 MHz
    & low      & none     & $\sim$0  & none \\
  \ppb \ra \bbb\ (\abseta $<$1) & 50 $\mu$b & 10 kHz
    & $\sim$6  & $\sim$1  & $\sim$0  & few \\
\hline
  \ppb \ra \Wboson $X$ \ra \lv $X$ & 4 nb & 0.8 Hz
    & high     & $\sim$45 & $\sim$45 & none \\
  \ppb \ra \Zboson $X$ \ra \bbb $X$ & 1 nb & 0.2 Hz
    & $\sim$45 & low      & $\sim$0  & $\sim$5 \\
\hline
  \ppb \ra \ttb \ra $\ell$+(\qrk{b})jets & 2.5 pb & 1.8/hour
    & $\sim$50 & $\sim$45 & $\sim$50 & $\sim$5 \\
  \ppb \ra \qrk{t}$X$ (s-chan) & 1 pb & 0.7/hour
    & & & & \\
  \ppb \ra \qrk{t}$X$ (t-chan) & 1 pb & 1.4/hour
    & & & & \\
\hline
  \ppb \ra \Wboson \Higgs \ra \lv \bbb & 26 fb & 0.4/day
    & $\sim$45 & $\sim$45 & $\sim$45 & $\sim$5 \\
  \ppb \ra \Zboson\ \Higgs \ra $\nu\nu$ \bbb & 22 fb & 0.4/day
    & $\sim$45 & none     & $\sim$70 & $\sim$5 \\
\hline
\hline
\end{tabular}
\end{center}
\end{table}

As mentioned before, production of
relatively low energy jets by QCD at these energies is a well understood
process (although other aspects of QCD observable at the Tevatron are
much more interesting). These events are therefore a major obstacle to
getting at the physics we don't understand, such as that associated
with the breaking of the electro-weak symmetry. The first step in
overcoming this obstacle is to write interesting events to tape for
offline analysis at a later stage. Obviously, it is impossible to do
this at the 10 MHz rate of QCD events. So sophisticated selection
mechanisms must be developed to winnow the few interesting events that
occur on a time scale of seconds to hours from the overwhelming QCD
background, all at a frequency set by the \ppb\ bunch crossing time of
396 ns. This daunting task is the job of the trigger system, which is
therefore one of the most critical elements in the experiments at the
Tevatron. 

\begin{figure}[ht]
\begin{center}
  \mygraphics{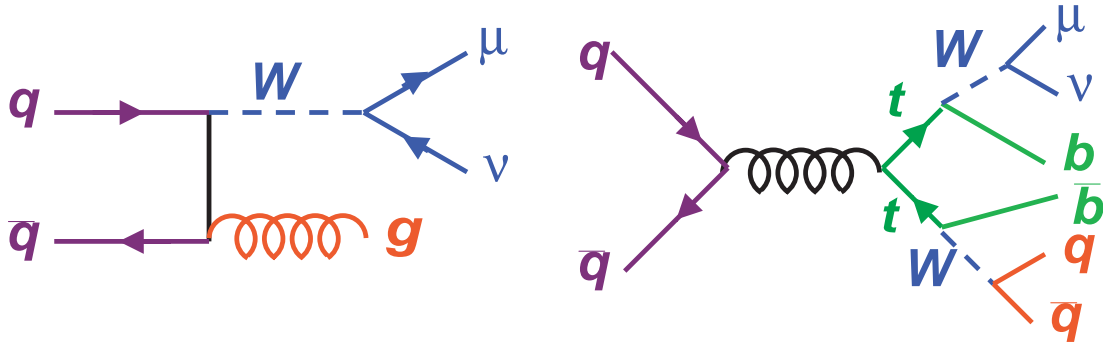}
  \caption{Feynman diagrams for \Wboson\ production with \Wboson \ra
    $\mu\nu$ (left-side) and top pair production with one top decaying
    semi-leptonically and the other decaying hadronically (right-side).}
  \label{fig:feynwt}
\end{center}
\end{figure}

\begin{figure}[h]
\begin{center}
  \mygraphics{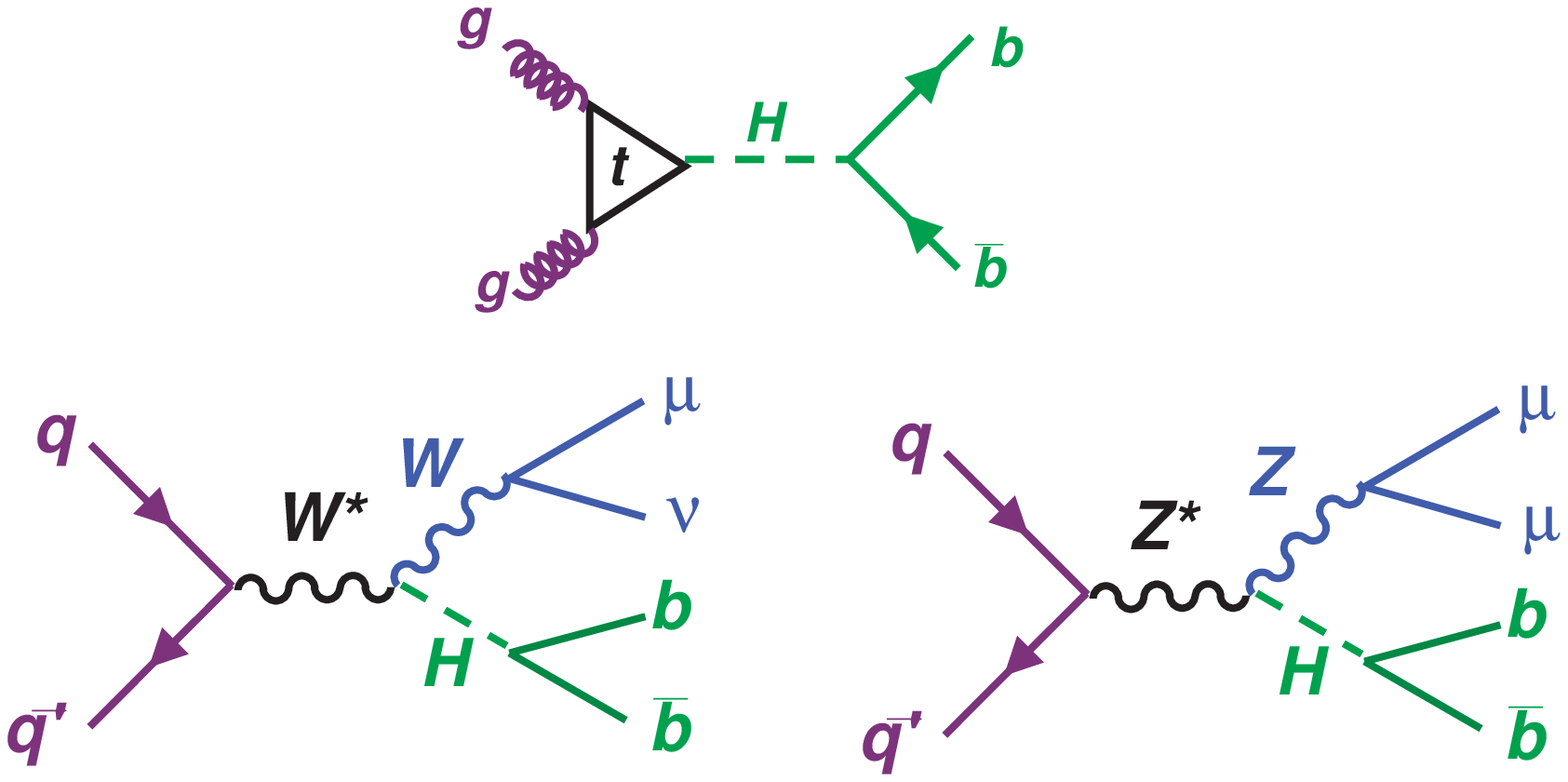}
  \caption{Feynman diagrams for Higgs production through gluon fusion
    (top diagram) and associated Higgs production with \Wboson 's and
    \Zboson 's (bottom diagrams).}
  \label{fig:feynhiggs}
\end{center}
\end{figure}

Offline, even more sophisticated algorithms are required to produce
clean samples of signal events with well-understood detector effects
and low background levels. This often requires choosing specific event
topologies for study, which also impacts the trigger algorithms
developed to select these events online.
For example, decays of vector bosons to quarks strongly resemble QCD
events, so only decays to leptons are generally used
(see Figure \ref{fig:feynwt} for a representative Feynman diagram).
The top quark decays nearly 100\% of the time via 
\qrk{t}\ra \qrk{b}\Wboson .
Again, leptonic decays of the \Wboson\ (for at least one of the tops
in the event) tend to give the cleanest event samples.
An example is given in Figure \ref{fig:feynwt}.
Finally, Higgs production at the Tevatron occurs mainly via
gluon-gluon fusion (through a top-quark loop) with the Higgs then
decaying either to \bbb\ (for Higgs mass less than about 130 \GeV ) or
to vector boson pairs (for higher Higgs masses).
While leptonic decays of the vector bosons 
be used to identify Higgs events in this production mode for high mass
Higgses, the \bbb\ final state seen in low mass Higgs production is
swamped by QCD produced \bbb\ pairs.
This forces us to search for a low mass Higgs in it associated
production mode
\begin{displaymath}
  \ppb \ra \Higgs + \Wboson (\Zboson )
  \ra \bbb + \lv , qq^{\prime} (\ell\ell, \nu\nu, qq)
\end{displaymath}
even though the cross-section for this is lower by almost an order of
magnitude than that for the gluon-gluon mode.
Feynman diagrams for low mass Higgs production are given in Figure
\ref{fig:feynhiggs}.

Luckily for trigger algorithm developers and offline analyzers, 
it turns out that these interesting physics channels share a
reasonably small number of simple event characteristics that allow them
to be distinguished from QCD
background. These characteristics all arise from two general features
of QCD events at 
hadron colliders contrasted to other (electro-weak) physics processes.

\begin{enumerate}
  \item The energy scales of the QCD hard scatter is set
        by the energy distribution of partons within the \hadron{p}
        and \antihad{p}, which are peaked at low values. This means
        that jets (and their component particles) in QCD events tend
        to have energies that are low compared to the beam energy. 
	In contrast, energy scales in the production of such objects
        as weak bosons, top, Higgs and beyond the SM particles are set
        by the heavy mass of the primary particle produced. Decay
        products of these particles then share their high energies.
	B-physics events tend to have energies intermediate between
        these two extremes as set by the \qrk{b}-quark mass.
  \item Hadronization of final-state partons from QCD hard scattering
        favors the production of low mass mesons and baryons
        containing light quarks (\qrk{u}, \qrk{d}, \qrk{s}). If
        unstable, these
        particles tend to have either very short 
        lifetimes (strong or EM) 
	or very long lifetimes. In addition, their low
        energy means that 
        they do not produce high \Pt\ leptons or neutrinos in their
        decays. 
	The heavy particles discussed above, however, can decay to
        high \Pt\ leptons or neutrinos. They also often have
        \qrk{b}-quarks in their decay chains. These quarks produce
        \hadron{B} hadrons with lifetimes such that they travel
        several mm's in the detector before decaying
	-- topologies that are reconstructible using precise tracking
        information from silicon detectors.
\end{enumerate}

Based on these differences, a short list of distinguishing variable
can be constructed that allow other physics processes to be
distinguished from QCD. These are shown below and typical values for
some of them are listed in Table \ref{table:tevphys}.

\begin{enumerate}
  \item Jet transverse energy -- \Etobj{jet}
  \item Lepton transverse energy -- \Etobj{lept}
  \item Missing transverse energy -- \MEt . 
	This can be caused by the presence of high \Pt\ neutrinos, or
	other non-interacting particles, that are not detected and
        therefore spoil the energy balance of the event in the
	transverse plane.
  \item Multi-particle vertices that are displaced from the point at
        which the \ppb\ interaction happened -- Displaced Vertices.
	Such vertices can arise
        from the decay of moderately long-lived
        \hadron{B}-hadrons that are produced at the interaction point.
  \item High energy photons 
\end{enumerate}

These variables, and others derived from them,
form the basis for most of the trigger and offline
algorithms used at CDF and \Dzero .

\section{First Results and a Look into the Future}
Having seen what the Run II capabilities are and generally how physics
is done at the Tevatron, we now turn our attention to the actual
CDF and \Dzero\ results. As mentioned before, only a small amount of
data had been analyzed at the time of the SLAC Summer Institute
(10--20 \invpb ) so the results discussed here, which are all
preliminary, give only a taste of
what can be done. We will, therefore, also examine what {\it can} be done
with Run II data set, with special attention paid to analyses to watch
in the coming years.

An enormous amount of work has gone into preparing these results and
predictions and justice certainly cannot be done to their beauty and
complexity in a few pages. Interested readers are encouraged to visit
the web sites of CDF\cite{cdfweb} and \Dzero \cite{d0web} for up to
the minute information about the experiments. The preliminary results
presented here were all prepared for the ICHEP conference in
Amsterdam. More details on them can be found in specific talks and
writeups (there were 21 from the Tevatron) linked off of the ICHEP02
web page \cite{ichep02}. 
Summaries were given by F. Bedeschi\cite{bedeschi} and
M. Narain\cite{narain}. 
Finally, predictions of CDF and \Dzero\
sensitivities were taken mainly from the Run II Tevatron Physics
Working Groups\cite{runiiWG} and also from some of the results
presented at the 2001 Snowmass workshop\cite{snowmass}.

One final note: B-physics at the Tevatron will not be discussed in the
following as it is extensively covered by F. W\"{u}rthwein in his
contribution to this conference\cite{wuerthwein} and in the Run II
B-Physics Working Group Report\cite{tevbwg}.

\subsection{QCD}
Although QCD events have been discussed mainly as
background to other physics processes many important studies of the
strong force will be performed at the Tevatron in Run II.
Among these are tests of (Next-to-)Next-to-Leading-Order, (N)NLO, QCD
using weak boson \Pt\ distributions and the angular distribution of
leptons from \Wboson\ decays.
Previous measurements of these distributions are statistics limited
and much more precise measurements should be possible in Run II. 
Direct photon production will also be used to test QCD and to measure
the gluon distribution in the proton where previous results are
inconsistent.
Searches for
deviations between data and predictions for well understood
QCD distributions, such as the di-jet invariant mass 
distribution, can also be used to detect evidence of physics beyond
the SM.
Finally, diffractive physics, such as studies of the properties of the
pomeron should also prove to be a rich field in Run II.

Aside from these important physics topics, QCD remains a large
background for many other analyses. It must therefore be thoroughly
understood before precise results can be obtained. Some of the issues
that will be addressed here in Run II are more precise tuning of Monte
Carlo event generators, better measurements of parton distribution
functions using \Wboson\ \Pt\ distributions, direct photon data and
high \Et\ jet data, and developing a better understanding of the
properties of various jet-finding algorithms.

\begin{figure}[h]
\begin{center}
\begin{minipage}[t]{0.475\textwidth}
  \begin{center}
    \mygraphics{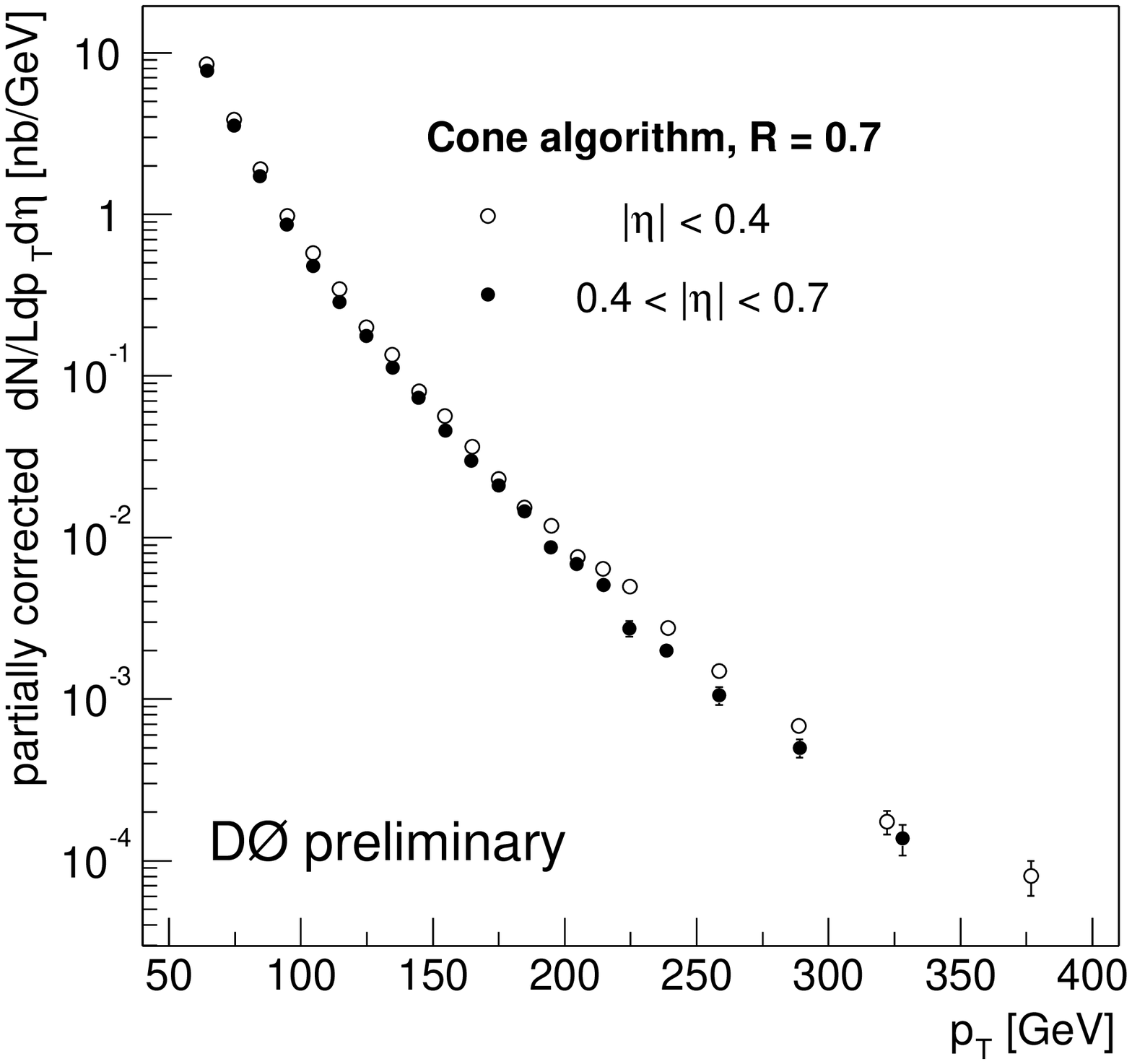}
  \end{center}
\end{minipage}
\hfill
\begin{minipage}[t]{0.475\textwidth}
  \begin{center}
    \mygraphics{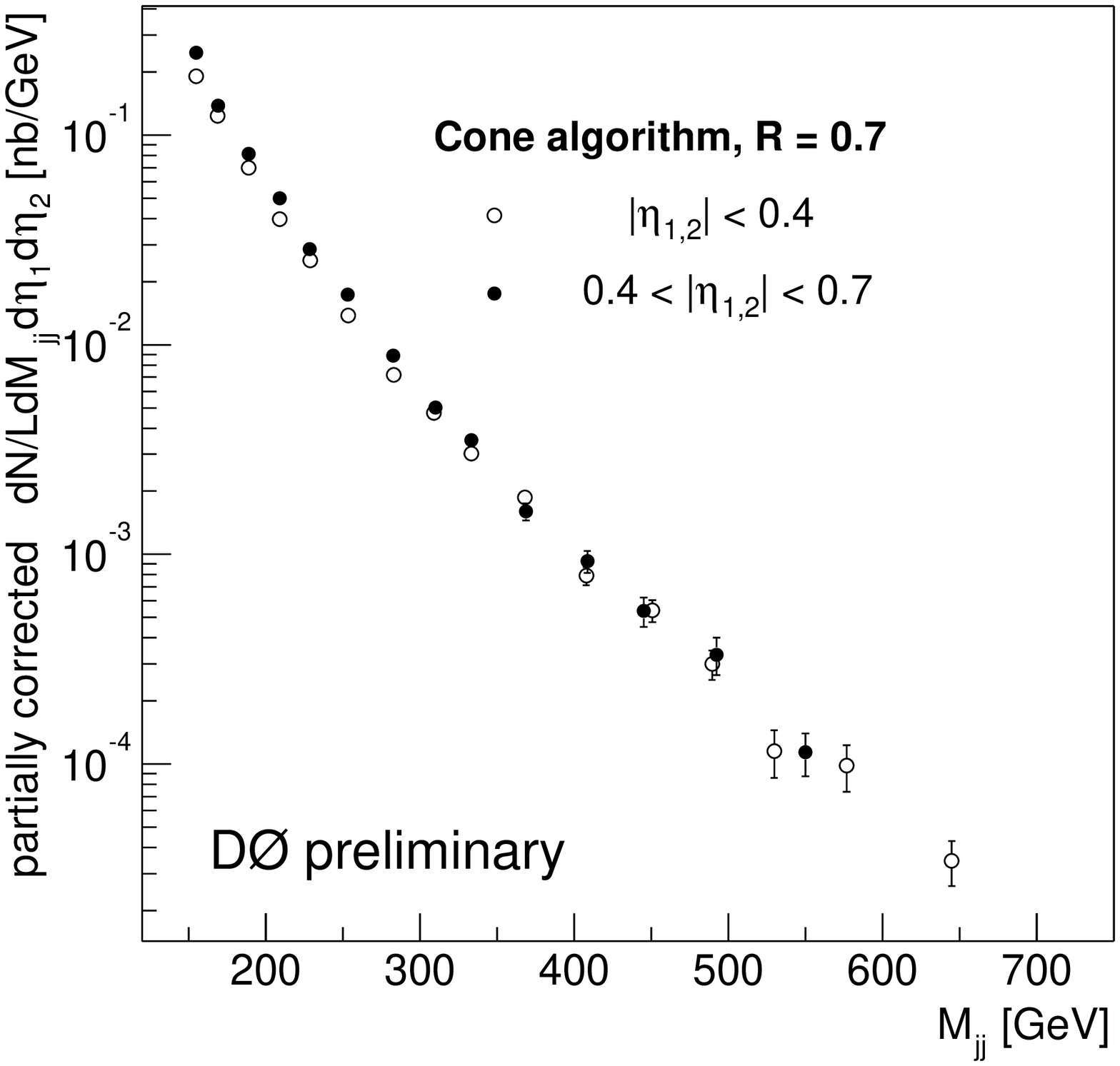}
  \end{center}
\end{minipage}
\caption{\Dzero\ preliminary inclusive jet \Pt\ distribution (left
  plot) and di-jet invariant mass spectrum (right plot).}
\label{fig:d0qcd}
\end{center}
\end{figure}

\begin{figure}[h]
\begin{center}
  \mygraphics{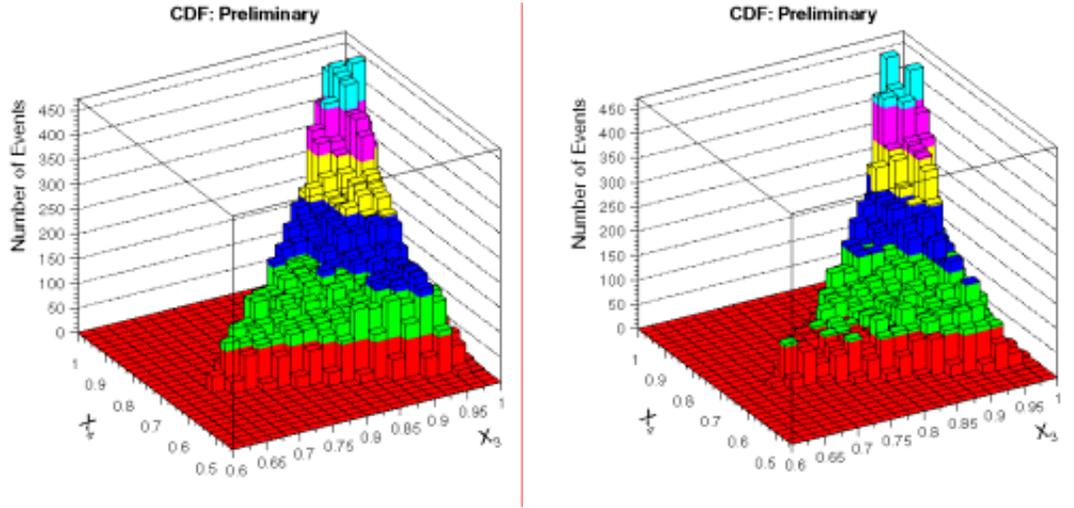}
  \caption{Preliminary Dalitz distributions of CDF 3-jet data (left
    plot) and NLO QCD prediction (right plot).}
  \label{fig:cdfqcd}
\end{center}
\end{figure}

Work has already begun on many of these topics and first physics
distributions were shown at the ICHEP02 conference\cite{dittman}.
As an example of the work shown,
\Dzero\ has produced preliminary jet \Pt\ spectra and di-jet invariant
mass spectra for events in two \abseta\ regions
(see Figure \ref{fig:d0qcd}). Although these distributions use
preliminary values for the jet energy scale and are not fully
corrected, they still show evidence of events with jet \Pt\ $>$ 400
GeV, which is an interesting region both for parton density
measurements and new physics searches.
CDF has also made great strides, showing the first comparison of
three-jet production with a 
NLO QCD prediction at a hadron collider.
Agreement between the data and the NLO QCD prediction for the Dalitz
variables $x_i = 2E_{jet-i}/m_{3-jet}$ is quite good
(see Figure \ref{fig:cdfqcd}).
The measured, total three-jet cross section in the kinematically
allowed region, 466$\pm$2$^{+206}_{-71}$ pb also agrees well with the
prediction of 402$\pm$3 pb.

\subsection{W/Z Boson and Top Physics}
Run II is expected to produce significant advances in our
understanding of the
properties of the \Wboson\ and \Zboson\ bosons and the top quark.
As can be seen from Table \ref{table:ewphys}, event yields for these
particles in Run II will be orders of magnitude more than those in Run
I. This will allow large improvements in precision on existing
measurements as well as opening up some new areas of study. A summary
of some of the most interesting electro-weak measurements in Run II,
compared to Run I results and expectations from the Large Hadron
Collider (LHC), are also given
in Table \ref{table:ewphys}.

\begin{table}
\begin{center}
\caption{Expected event yields and measurement precisions for various
  \Wboson , \Zboson\ and \qrk{t}-quark observables compared the
  current status and predictions from the LHC.}
\label{table:ewphys}
\begin{tabular}{|c|c|cc|c|}
\hline
\hline
  {\bf Measurement} & {\bf Current} 
  & \multicolumn{2}{|c|}{\bf Run II / exp.} & LHC \\
  & & (2 \invfb ) & (15 \invfb ) & (10 \invfb ) \\
\hline
\hline
  reconstr. \Wboson \ra \lv
    & 77k & 2300k & 17250k & \scinot{6}{7} \\
  reconstr. \Zboson \ra \lplm
    & 10k & 202k  & 1515k  & \scinot{6}{6} \\
  reconstr. \ppb \ra \ttb \ra $\ell$+jets
    & $\sim$20 & $\sim$800  & $\sim$6000  
    & \scinot{8}{5} \\
  reconstr. \ppb \ra \qrk{t}$X$
    & 0        & $\sim$150  & $\sim$1200  
    & \scinot{1.7}{4} \\
  & (Ref. \citenum{glenzinski}) 
    & \multicolumn{2}{|c|}{(Ref. \citenum{glenzinski})}
    & (Ref's \citenum{lhcew,lhctop}) \\
\hline
  $\delta \sin^2 \theta_W$
    & \scinot{5.1}{-4}
    & & \scinot{4}{-4} 
    & \scinot{1.4}{-4} (100 \invfb ) \\
  & (Ref. \citenum{lepewwg}) 
    & \multicolumn{2}{|c|}{(Ref. \citenum{tevewwg})}
    & (Ref. \citenum{tevewwg}) \\
\hline
  $\delta$\MW [MeV] & 39 
    & 27 & 17 & 10 \\
  $\delta$\Mt [GeV] & 5.1
    & 2.7 & 1.3 & $<$2 \\
  & (Ref. \citenum{pdg})
    & \multicolumn{2}{|c|}{(Ref. \citenum{snowew})}
    & (Ref's \citenum{lhcew,lhctop}) \\
  $\delta$\MH /\MH [\%] & 58 & 35 & 25 & 18 \\
  (Ref. \citenum{snowew}) &  &    &    & \\
\hline
  $\delta BR(t \ra W_o b)$ [\%]  & 42 
    & 9 & 4 & 1.6 \\
  & (Ref. \citenum{cdftpol})
    & \multicolumn{2}{|c|}{(Ref. \citenum{iashvilli})}
    & (Ref. \citenum{lhctop}) \\
\hline
  $\sigma (\ppb \ra tX)$         & $<$13.5 pb 
    & 20\% & 8\% & \\
  \abs{V_{tb}}                   & ---        
    & 12\% & 5\% & $<$5\% (100 \invfb ) \\
  & (Ref. \citenum{cdf1top}) 
    & \multicolumn{2}{|c|}{(Ref. \citenum{iashvilli})} 
    & (Ref. \citenum{lhctop}) \\
\hline
\hline
\end{tabular}
\end{center}
\end{table}

\begin{figure}[h]
\begin{center}
  \mygraphics{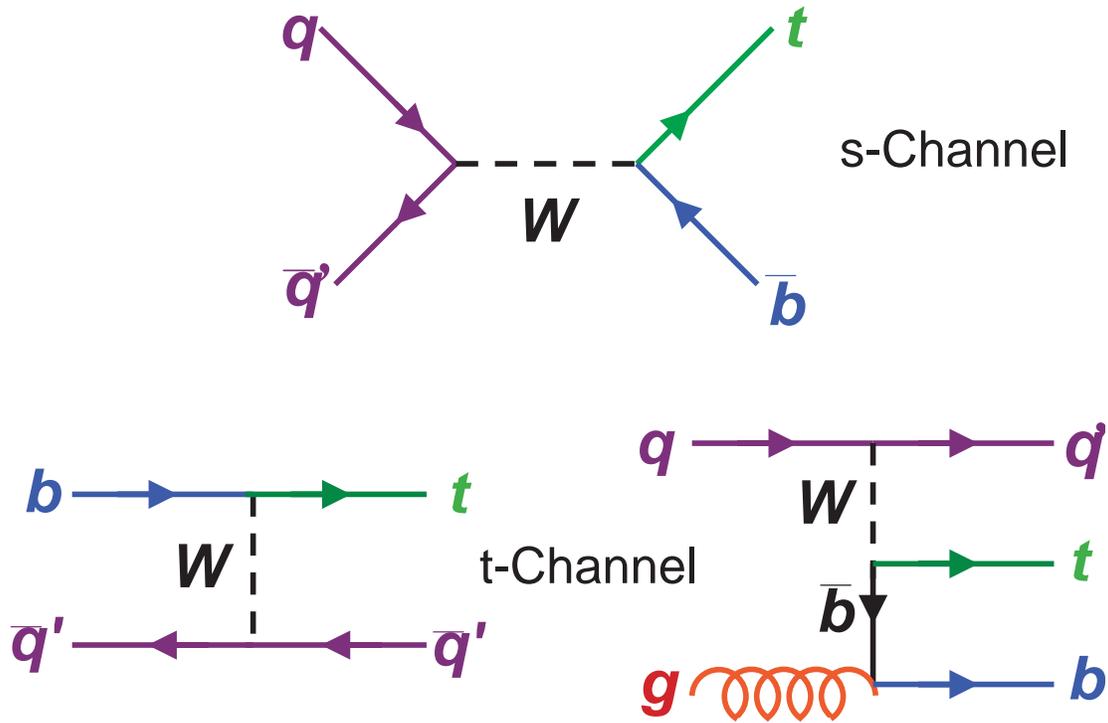}
  \caption{Feynman diagrams for single top production in the $s$- and
    $t$-channels.}
  \label{fig:singlet}
\end{center}
\end{figure}

Several topics deserve specific note. 
Measurement of the forward-backward asymmetry in \Zboson \ra
\lplm\ events with the Run IIb data set
will allow a determination of the $\sin^2
\theta_W$ to a precision comparable to the current world average.
The precision on the mass
measurement of the \Wboson\ and the top quark, for each experiment, will
be improved to 
approximately 30 MeV and 3 GeV respectively in Run IIa. 
These variables are
sensitive to the Higgs mass because of loop corrections. Comparing the
Higgs mass value predicted from \MW , \Mt\ and other electro-weak
variables to the value that is found if the Higgs is observed will be
a stringent test of whether the observed Higgs is as predicted in the
SM.
Measuring the decays of the top to a longitudinally polarized
\Wboson\ and a \qrk{b}-quark will provide a test of the $V-A$
structure of the weak interaction in the top sector.
Finally, a new area that will open up in Run II will be single-top production
through \Wboson -boson exchange in the $s$- and $t$-channels. 
Feynman diagrams for these processes are shown in Figure
\ref{fig:singlet}.
This
will allow, for the first time, a direct measurement of the CKM matrix
element \abs{V_{tb}}.
Predictions for the precision of these measurements compared to
current and future values are given in Table \ref{table:ewphys}.

\begin{figure}[h]
\begin{minipage}[t]{0.475\textwidth}
  \begin{center}
    \mygraphics{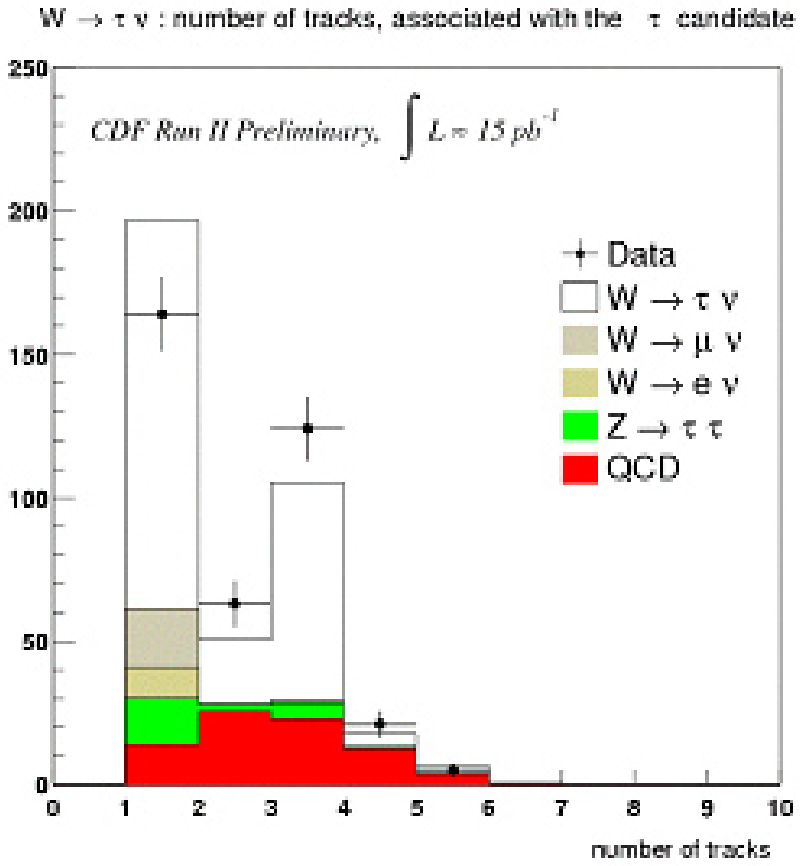}
    \caption{Track multiplicity in the $\tau$-candidate jet for the
      CDF \Wboson \ra \tauv\ selection.}
    \label{fig:cdftau}
  \end{center}
\end{minipage}
\hfill
\begin{minipage}[t]{0.475\textwidth}
  \begin{center}
    \mygraphics{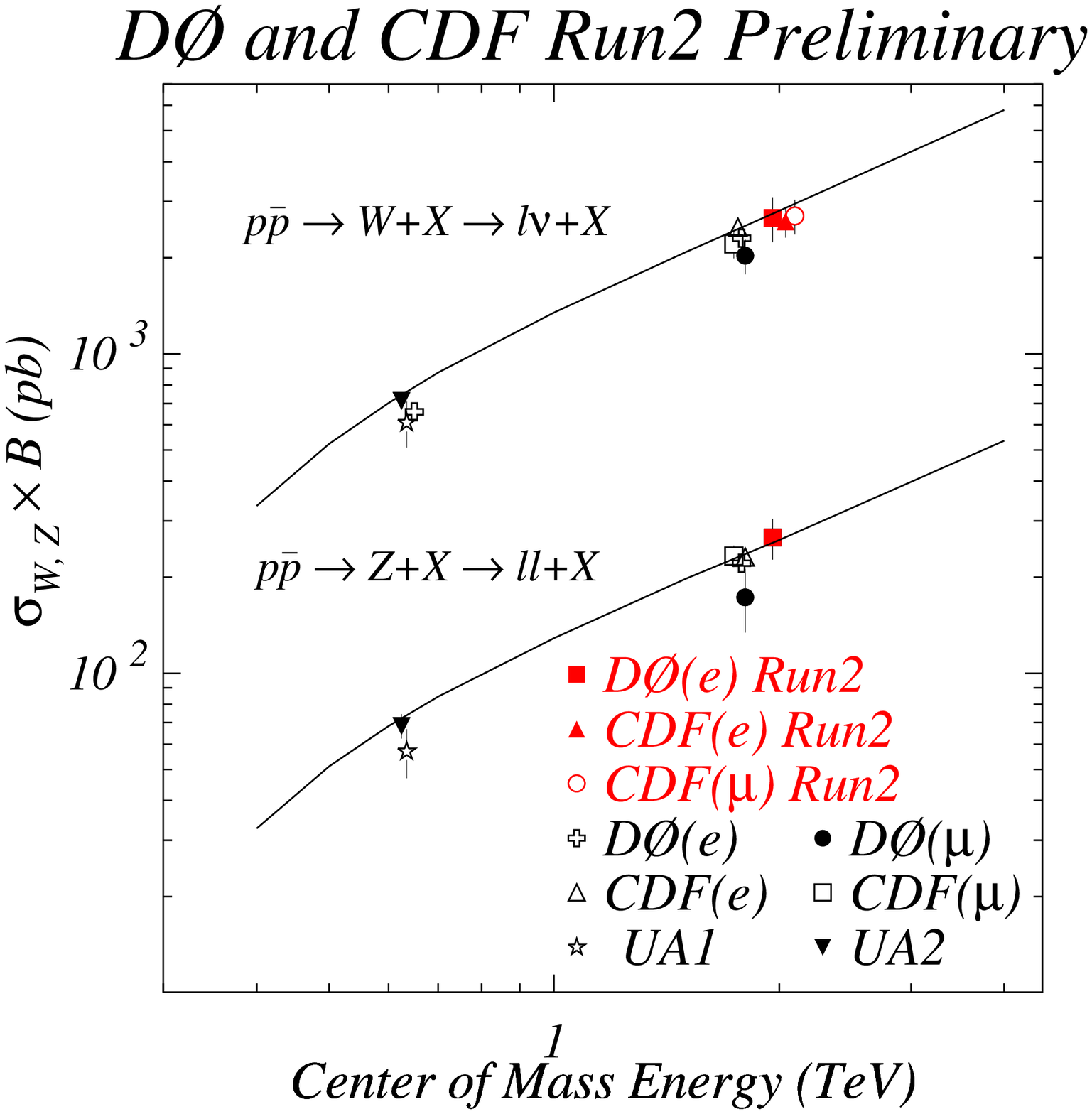}
    \caption{A comparison of the Run II CDF and \Dzero\ measurements of the
      \Wboson\ and \Zboson\ production cross-sections times leptonic
      branching ratios at \sqrts =1.96\TeV\ with previous measurements
      and theoretical predictions.}
    \label{fig:wzxs}
  \end{center}
\end{minipage}
\end{figure}

CDF and \Dzero\ have produced first results on the road to these
exciting measurements\cite{glenzinski,iashvilli}. 
Hundreds of \Zboson \ra \lplm\
and thousands of \Wboson \ra \lv\ events have been observed (see
Table \ref{table:ewmeas}) and CDF has even observed a \Wboson \ra
\tauv\ signal as an excess of one- and three-track jets (most $\tau$
decays have one or three charged particles) in events with narrow jets
consistent with \Wboson \ra \tauv\ production (see Figure
\ref{fig:cdftau}). First measurements of cross-section times branching
ratio have also been made for both \Wboson\ and \Zboson\ production in
the electron and muon channels as shown in Table \ref{table:ewmeas}
and Figure \ref{fig:wzxs}. These measurements are quite consistent
with theoretical predictions and show clearly the evolution of the
cross-section with energy when compared with previous measurements.

A preliminary measurement of the \Wboson\ width has also been made by
both collaborations using the ratio of the \Wboson\ to \Zboson\
cross-section times branching ratio measurements
(see Table \ref{table:ewmeas}).
This technique relies on
input from LEP for the \Zboson \ra \epem\ branching ratio
and from theory for a prediction of the \Wboson /\Zboson\
cross-section ratio and is not the ultimate method that will be used
to determine this quantity using Run II data.
However the good
agreement with the theoretical prediction is an indication that we are
on the right track to making competitive measurements using \Wboson 's
and \Zboson 's.

\begin{table}
\begin{center}
\caption{Numbers of candidate \Wboson\ and \Zboson\ events in the
  electron and muon decay channels observed by CDF and \Dzero . Also
  shown are preliminary measurements of \Wboson\ and \Zboson\
  production cross-section times branching ratio measurements with
  errors from event statistics, systematics and the luminosity
  measurement, in that order.}
\label{table:ewmeas}
\begin{tabular}{|c|cc|cc|}
\hline
\hline
  {\bf Channel} & \multicolumn{2}{|c|}{\bf Candidates} 
  & \multicolumn{2}{|c|}{\boldmath$\sigma \times BR$ [pb]} \\
  & {\bf CDF} & {\bf \Dzero } & {\bf CDF} & {\bf \Dzero } \\
\hline
\hline
  \Wboson \ra \ev     & 5547 & 9205
    & 2600$\pm$30$\pm$130$\pm$260 & 2670$\pm$60$\pm$330$\pm$270 \\
  \Zboson \ra \epem   & 798  & 328
    & --- & 266$\pm$20$\pm$20$\pm$27 \\
\hline
  \Wboson \ra \muv    & 4561 & ---
    & 2700$\pm$40$\pm$190$\pm$270 & --- \\
  \Zboson \ra \mupmum & $\sim$170  & $\sim$57
    & --- & --- \\
\hline
  $\Gamma_W$ [\GeV ] 
    & \multicolumn{2}{|c|}{2.118$\pm$0.042\cite{pdg}}
    & 1.67$\pm$0.24$\pm$0.14 
    & 2.26$\pm$0.18$\pm$0.29$\pm$0.04 \\
  & \multicolumn{2}{|c|}{(world ave)} 
    & ($\pm$stat,syst) & ($\pm$stat,syst,theory) \\
\hline
\hline
\end{tabular}
\end{center}
\end{table}

\begin{figure}[h]
\begin{minipage}[t]{0.475\textwidth}
  \begin{center}
    \mygraphics{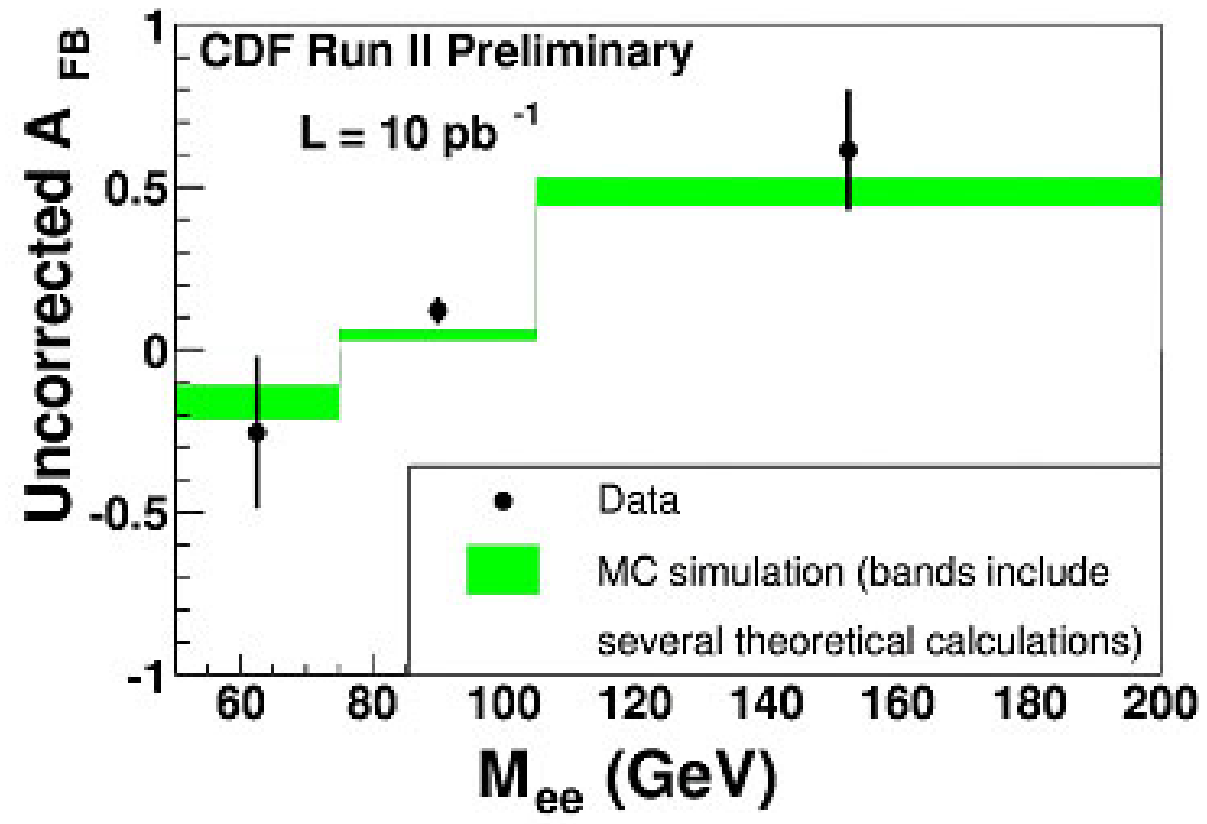}
    \caption{The forward-backward asymmetry as a function of \epem\
      invariant mass for \Zboson \ra \epem\ events from CDF.}
    \label{fig:cdfafb}
  \end{center}
\end{minipage}
\hfill
\begin{minipage}[t]{0.475\textwidth}
  \begin{center}
    \mygraphics{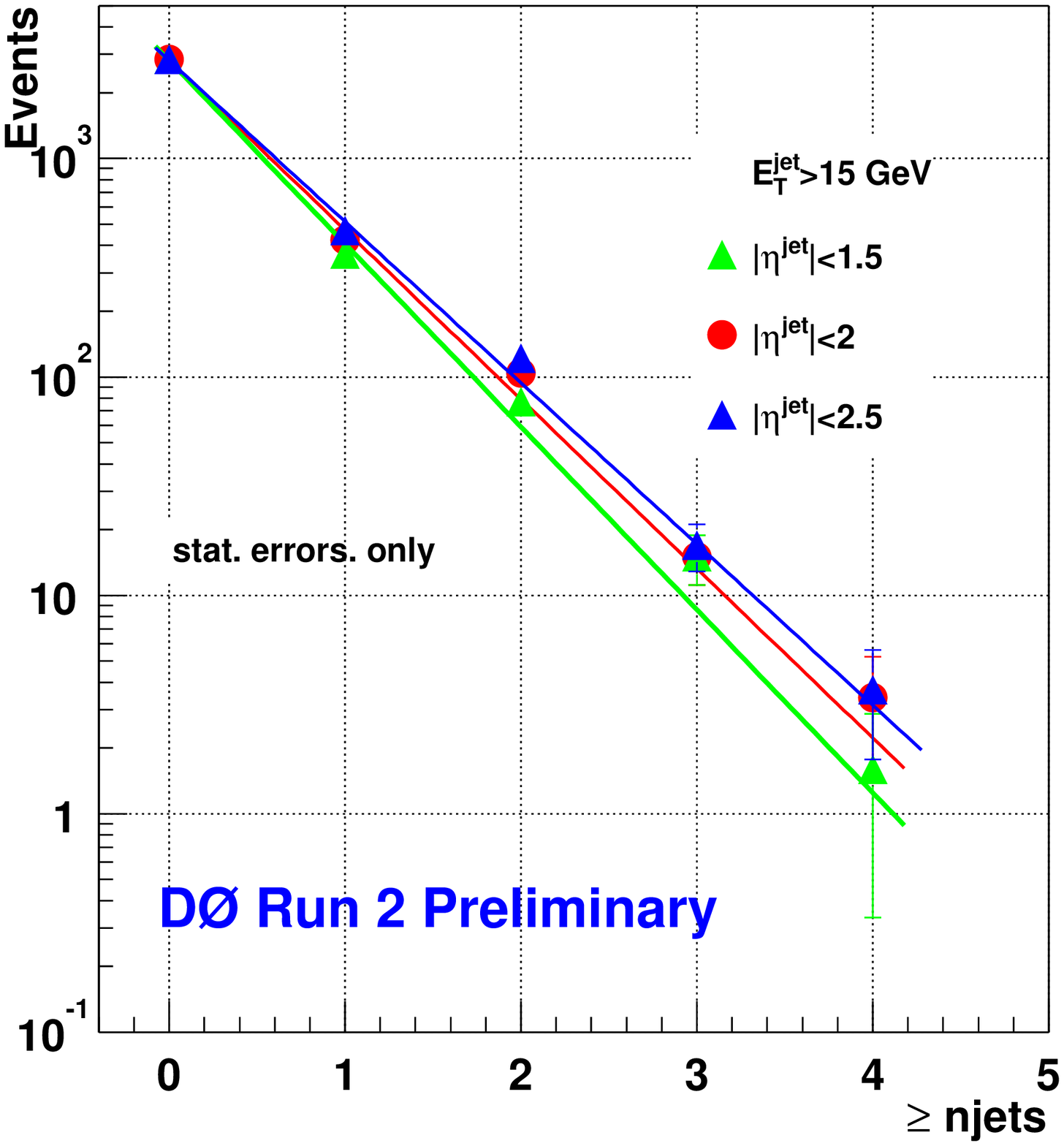}
    \caption{The \Dzero\ measurement of \Wboson (\ra \ev )+jets
      vs. number of reconstructed jets.}
    \label{fig:d0wjet}
  \end{center}
\end{minipage}
\end{figure}

Preliminary results for other production properties of vector bosons
are also being made. CDF has produced a first distribution of the
forward-backward asymmetry vs \epem\ invariant mass for \Zboson \ra
\epem\ events. Remember, that this can be used to determine 
$\sin^2 \theta_W$. Although statistics are still limited, the data
agrees quite well with theoretical expectations as can be seen in
Figure \ref{fig:cdfafb}.

On the top quark side, \Dzero\ has begun the process by examining
\Wboson +jets events. These are interesting for many analyses, with
top events inhabiting the \Wboson +$\geq$3-jets sample and Higgs
possibly showing up in \Wboson /\Zboson+$\geq$2-jets. Although
statistics are still to low to have observed any top events (let alone
the elusive Higgs) the \Dzero\ data shown in Figure \ref{fig:d0wjet}
indicate that we have made a strong start.

\subsection{The Higgs}
\begin{figure}[h]
\begin{minipage}[t]{0.475\textwidth}
  \begin{center}
    \mygraphics{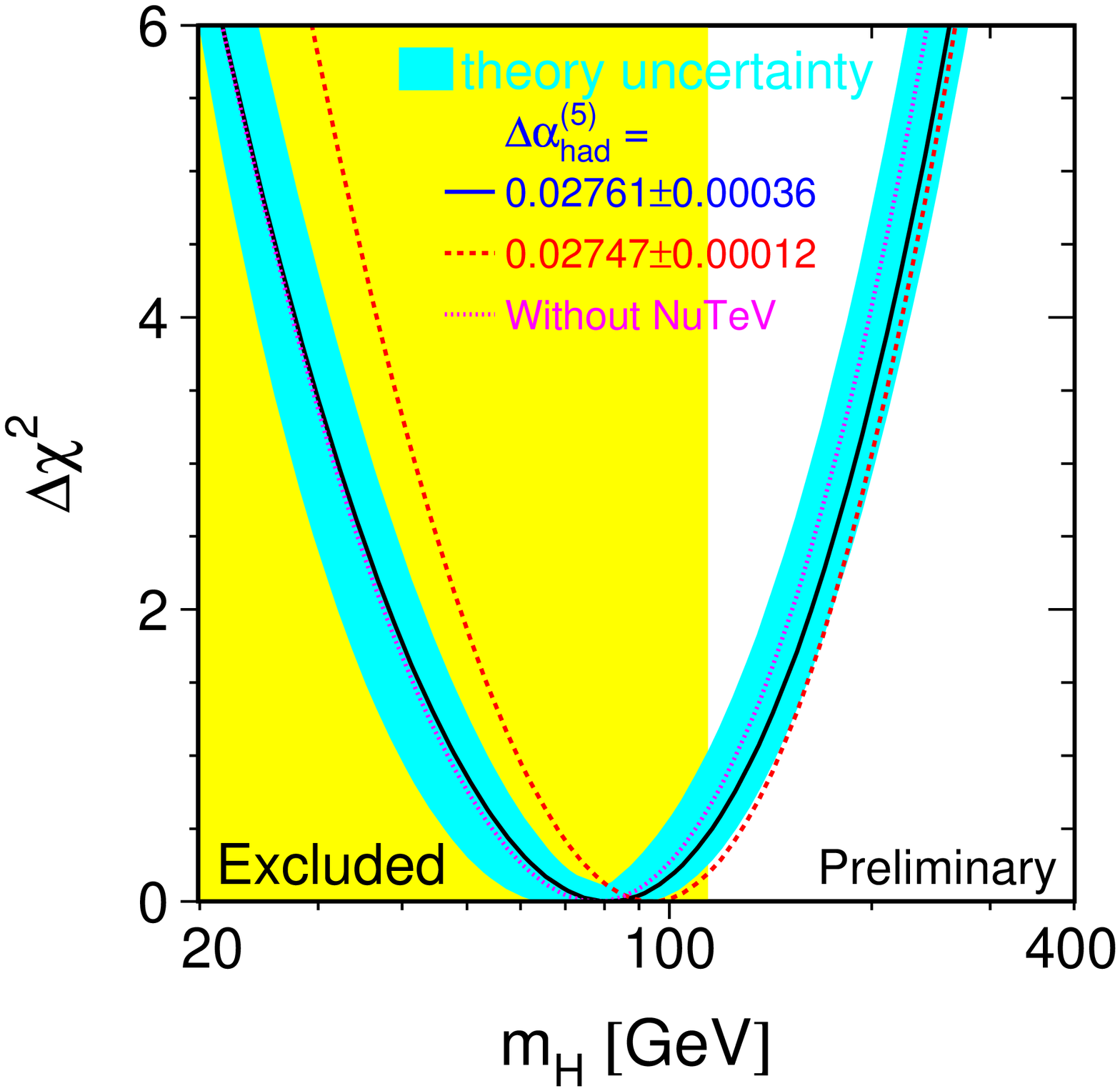}
    \caption{$\Delta \chi^2$ curve for the fit of the Higgs mass to
      precision electro-weak data\cite{lepewwg}.}
    \label{fig:lepewhiggs}
  \end{center}
\end{minipage}
\hfill
\begin{minipage}[t]{0.475\textwidth}
  \begin{center}
    \mygraphics{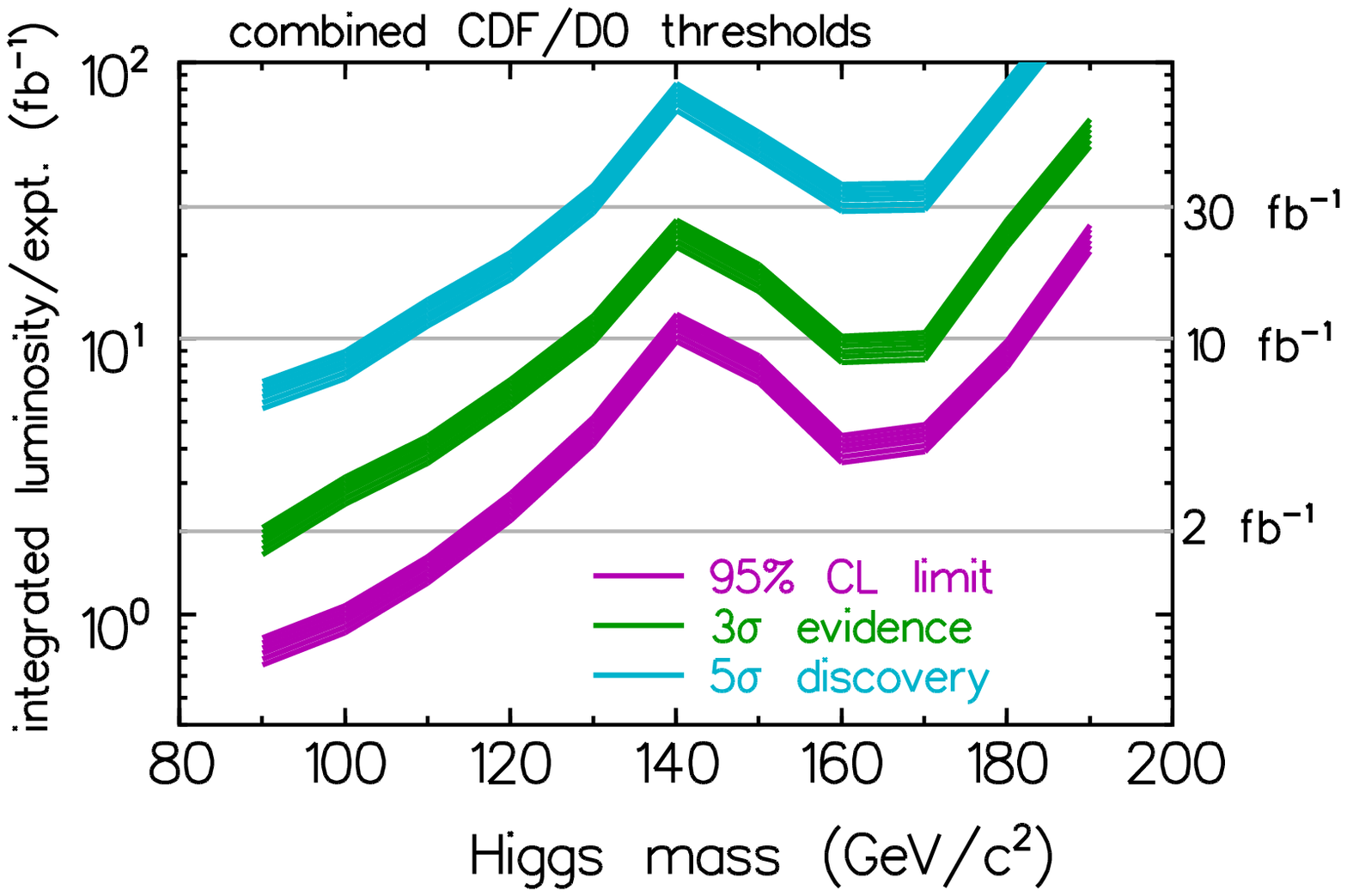}
    \caption{Predictions of Higgs sensitivity vs. mass for Run
      II\cite{tevhiggswg}.}
    \label{fig:tevhiggs}
  \end{center}
\end{minipage}
\end{figure}

One of the most exciting prospects for Run II at the Tevatron is the
possibility of discovering the Higgs. 
The opportunity is evident from an examination of Figures
\ref{fig:lepewhiggs} and \ref{fig:tevhiggs}.
Figure \ref{fig:lepewhiggs}, which is the LEP Electroweak Working
Group fit of the Higgs mass to precision electro-weak
observables\cite{lepewwg}, indicates that, if the Higgs is SM-like, it
should be light. The fit gives \MH\ = 81$^{+52}_{-33}$ \GeV , or
\MH\ $<$ 193 \GeV\ at 95\% CL,
which is compatible with
direct searches at LEP II\cite{lephiggs} that limit this range to
\MH\ $>$ 114.4 \GeV\ at 95\% CL.
A light Higgs is also required by the minimal supersymmetric extension
to the SM, which restricts the lightest Higgs mass to be less than
approximately 135 \GeV .
All of this is good news for the Tevatron, as can be seen from Figure
\ref{fig:tevhiggs}. Given the luminosities hoped for in Run IIb, we
should be able to see 3$\sigma$ evidence for an SM-like Higgs up to
masses of about 180 \GeV .

To accomplish this feat, however, information from all Higgs
production and decay modes must be used.
\begin{itemize}
  \item \MH\ $<$ 130 \GeV : 
    \qrk{q}\aqrk{q}$^{\prime}$ \ra\ \Wboson /\Zboson\ \Higgs\ \ra\
    \lv \bbb ,\vvb \bbb ,\lplm \bbb ,\qqb \bbb
  \item \MH\ $>$ 130 \GeV :
    \gluon \gluon \ra\ \Higgs\ \ra\ $W^+ W^-$/\Zboson /\Zboson\
    \ra\ \lplm \vvb ,\lplm jj, \lplm $\ell^{\prime +}$
\end{itemize}
Nearly all aspects of the detectors will be used in these searches.
However, triggers, particularly those involving leptons and jets+\MEt ,
will need to be nearly 100\% efficiency for Higgs events,
and \qrk{b}-quark identification will have to be superb --
with tagging efficiencies of 60--75\% and \bbb\ resonance mass
resolution of $\sim$30\% required.

\begin{figure}[h]
\begin{center}
\begin{minipage}[t]{0.475\textwidth}
  \begin{center}
    \mygraphics{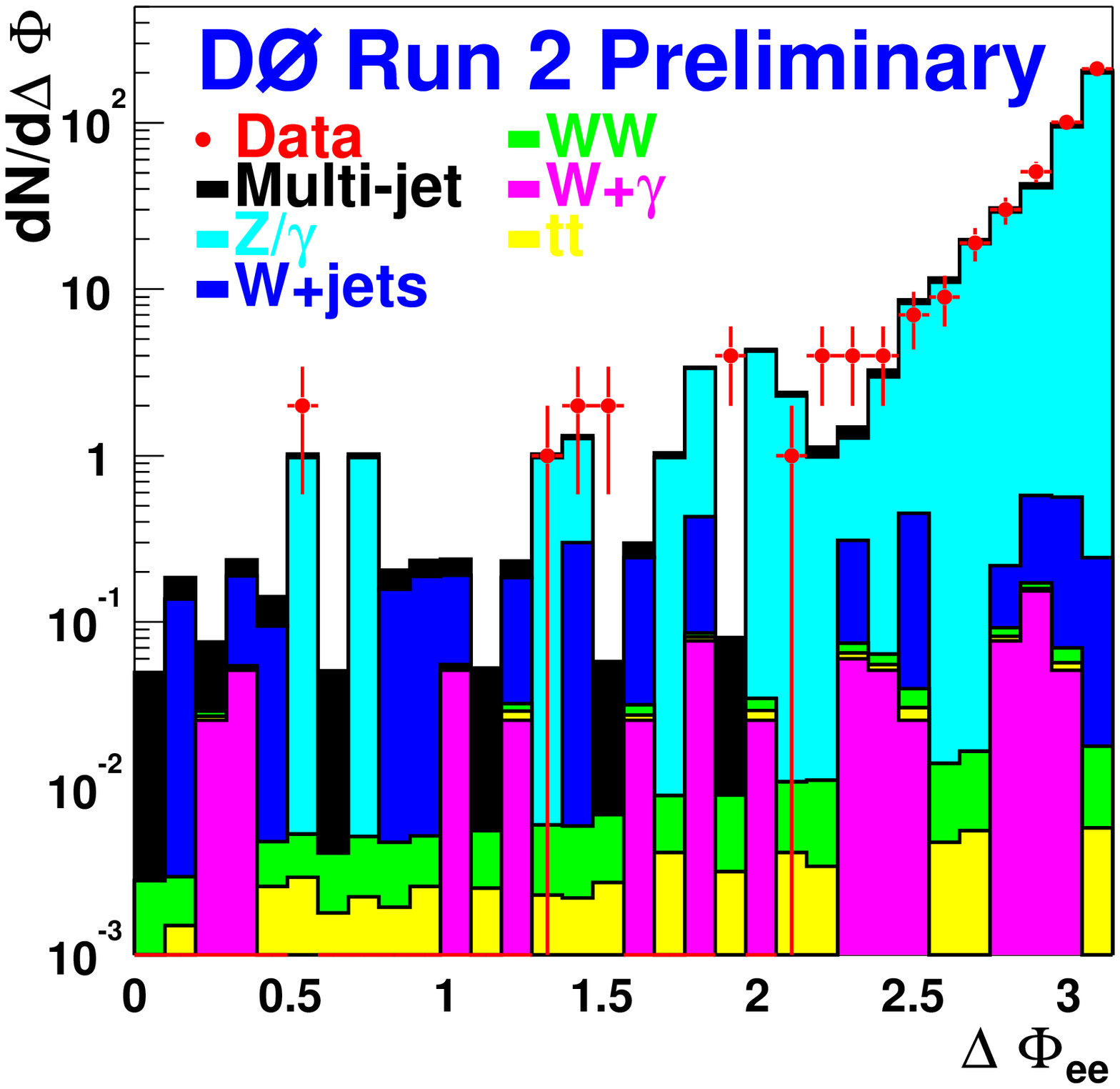}
  \end{center}
\end{minipage}
\hfill
\begin{minipage}[t]{0.475\textwidth}
  \begin{center}
    \mygraphics{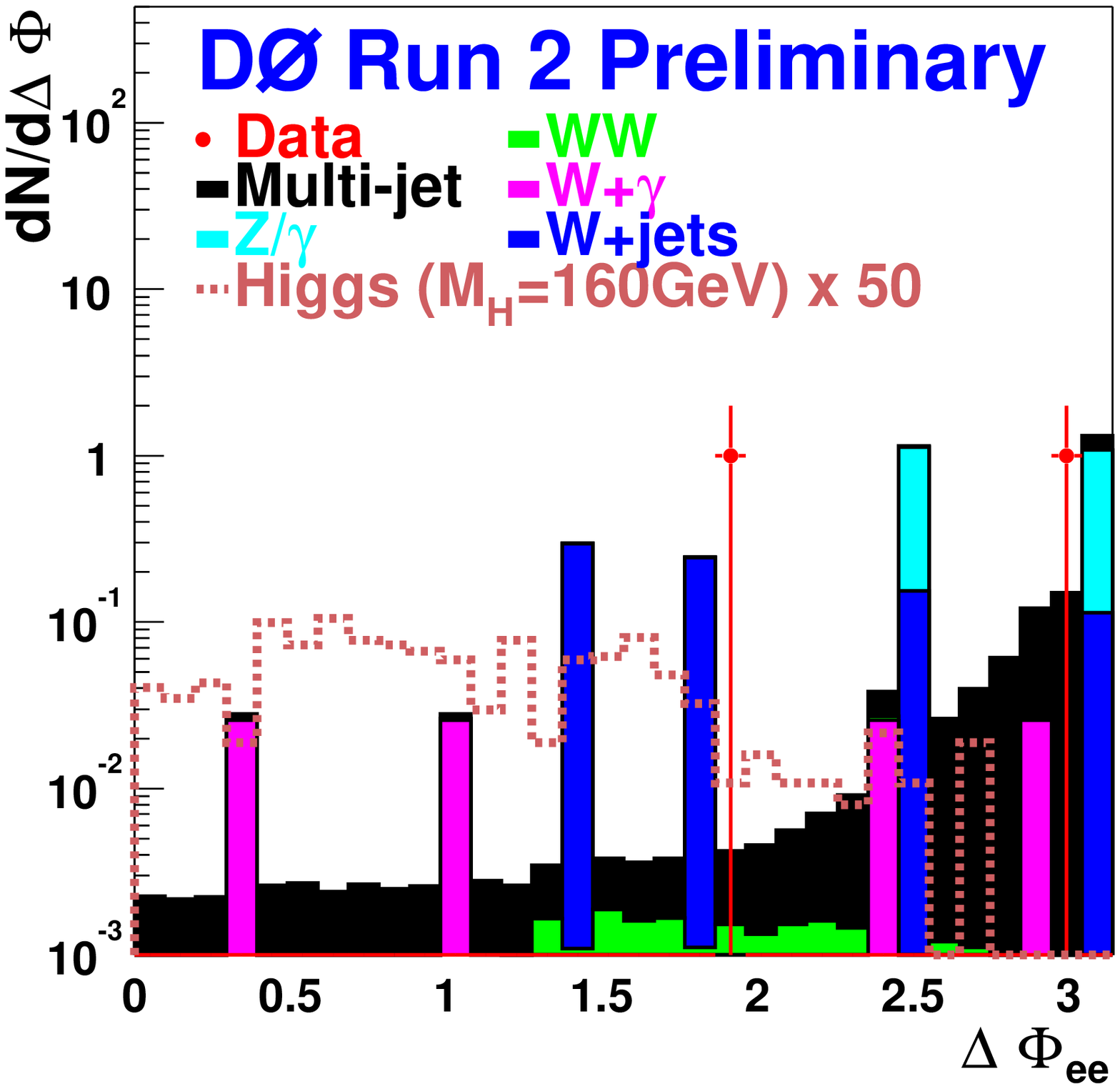}
  \end{center}
\end{minipage}
\caption{The \epem\ azimuthal opening angle in the \Dzero\
  \Higgs \ra \Wboson \Wboson \ra \epem \vvb\ search.
  The left plot shows data containing good \epem\ pairs while the
  right plot shows the distribution after all selection cuts (large
  \MEt , no jets, etc.).}
\label{fig:d0higgs}
\end{center}
\end{figure}

The game is particularly intense because the LHC experiments should be
able to see nearly any type of Higgs within a year of their
start. This has motivated CDF and \Dzero\ to start looking at Higgs signatures
even though several years worth of accumulated luminosity will be
required before even the LEP limits can be passed.
\Dzero\ has started the process by looking for events containing
\epem\ and \MEt\ (Ref's \citenum{glenzinski,iashvilli}). 
Excesses of such events over SM backgrounds are
predicted in models where the Higgs coupling to fermions is
suppressed. In these cases the branching ratio of 
\Higgs \ra $W^+ W^-$ is $\sim$98\% for \MH $>$100\GeV .
Preliminary results from \Dzero\ are shown in Figure
\ref{fig:d0higgs}.
No excess above predicted backgrounds is observed with the 9 \invpb\
analyzed.
Even though no signal is observed this study is important in that it
increases confidence in our ability to model backgrounds to Higgs
production. 

\subsection{Beyond the Standard Model}
The limitations of the SM lead most physicists to believe that it
cannot be the whole story behind the fundamental forces of nature. 
Unfortunately, no evidence for any chink in the armor of the SM has
yet been found, although a bewildering array of theoretical models
have been proposed to address the SM's shortcomings 
(see Ref's \citenum{tevhiggswg}-\citenum{tevgmsusywg}).
Because of the excellent detector capabilities and the large data set,
Run II should be a fruitful field in which to search for signals of
physics beyond the SM. 

Sensitivities for a sample of new physics topics are shown in Table
\ref{table:newphen}. While this list shows only the tip of the new
physics iceberg, it
does give a sense of the capabilities of the Tevatron in
beyond the SM searches. A few comments about the entries in the list
are in order.

\begin{table}
\begin{center}
\caption{A selection of new physics sensitivities at the Tevatron in
  Run II compared to the current state of knowledge and LHC
  expectations.}
\label{table:newphen}
\begin{tabular}{|c|c|c|c|}
\hline
\hline
  & & \multicolumn{2}{|c|}{\bf Sensitivity} \\
  {\bf Model} & {\bf Current Excl.} & {\bf Run IIb}
    & {\bf LHC} (100\invfb ) \\ 
\hline
\hline
  MSSM 
    & 0.5$< \tan \beta <$2.4 
    & $\sim$all 5$\sigma$ 
    & all \\
  ($\tan \beta , M_A$) 
    & (Ref. \citenum{lepsusy})
    & (Ref. \citenum{tevhiggswg})
    & \\
\hline
  Technicolor 
    & $>$600 
    & $>$850 
    & $\gg$800 \\
  ($M_{\rho T8}$ [\GeV ])
    & (Ref. \citenum{orejudos})
    & (2\invfb , Ref. \citenum{orejudos})
    & (Ref. \citenum{atlastdr}) \\
\hline
  Extra Dim. 
    & $>$1.0--1.4 
    & 2.1--3.5 
    & 6-9 \\
  (\MS\ [TeV ])
    & (Ref. \citenum{orejudos})
    & (Ref. \citenum{ledpred})
    & (Ref. \citenum{tovey}) \\
\hline
  Rare Top (\qrk{t}\ra \qrk{q}\Zboson )
    & $<$33\% 
    & \scinot{2}{-3} 
    & \scinot{2}{-4} \\
  (\qrk{t}\ra \qrk{q}$\gamma$)
    & $<$3.2\% 
    & \scinot{2}{-4} 
    & \scinot{3.4}{-5} \\
    & (Ref. \citenum{cdfraret})
    & (Ref. \citenum{iashvilli})
    & (Ref. \citenum{lhctop}) \\
\hline
  New bosons 
    & $>$690 
    & 900--1200 
    & 5000 \\
  ($M_{Z\prime}$ [\GeV ])
    & (Ref. \citenum{orejudos})
    & (Ref. \citenum{orejudos})
    & (Ref. \citenum{lcexotic}) \\
\hline
\hline
\end{tabular}
\end{center}
\end{table}

Given the {\it ad hoc} nature of electro-weak symmetry breaking in the
SM, several theoretical frameworks have been developed to attempt to
explain it.
One example is {\it supersymmetry}, or {\it SUSY}
\cite{susy}, which  postulates superpartners
having particles with spins differing by one half a unit
from all known SM particles
as well as an extended Higgs sector containing
five physical Higgs states.
In its raw state, SUSY contains more than 100 parameters allowing it
to predict almost any type of new phenomenon that could be
observed. This situation can be improved (from an experimentalist's
point of view) by making various, well motivated, assumptions about
relationships between parameters. Different types of assumptions lead
to different versions of SUSY, one of the most popular of which is the
Minimal Supersymmetric Standard Model, or MSSM, where only five
parameters are required to described the model.
(See Ref. \citenum{tevhiggswg} for a good overview.)

Another possibility is {\it technicolor} \cite{techni}, which
breaks electroweak symmetry when the interactions of an extended
gauge symmetry containing extra multiplets of technifermions become
strong.
Still another option lies in theories with large extra
dimensions\cite{ledorig},
which solves the problem of the disparity between the
Planck scale at which gravity is strong and the electro-weak scale by
assuming that the natural energy scale of gravity, (\MS ) 
actually {\it is} near the
electro-weak scale. This assumption is made plausible by postulating
the existence of extra spatial dimensions in which gravity also acts,
and leads to the prediction of a number of new graviton states with
various couplings to fermions and bosons.
Other theories predict an enhancement in the rate of decays of the top
quark that 
are highly suppressed in the SM\cite{raretop} or the existence of
heavier versions of the \Wboson - and \Zboson -bosons\cite{zprime}.

As can be seen from Table \ref{table:newphen}, the precision with
which all of these theories (and many more) can be tested will be
substantially improved in Run II at the Tevatron.

\begin{figure}[h]
\begin{minipage}[t]{0.475\textwidth}
  \begin{center}
    \mygraphics{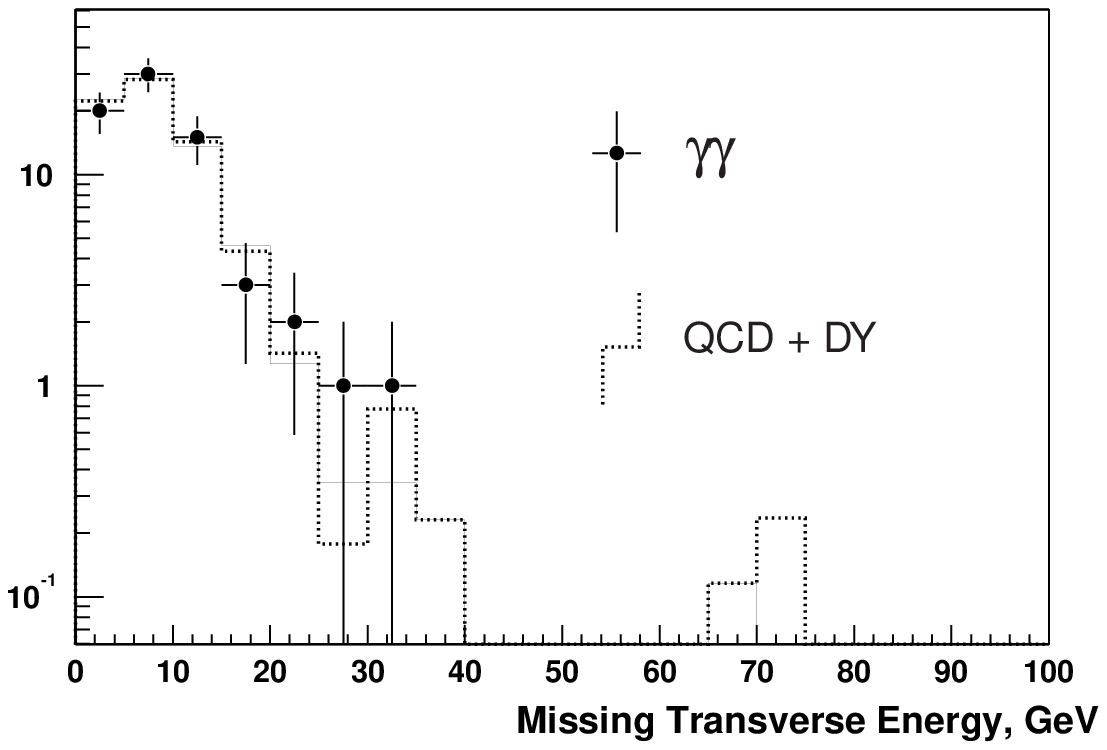}
    \caption{Preliminary \Dzero\ results for a gauge mediated SUSY
      search. Show is the distribution of missing \Et\ in 
      $\gamma \gamma$ events.}
    \label{fig:d0gmsusy}
  \end{center}
\end{minipage}
\hfill
\begin{minipage}[t]{0.475\textwidth}
  \begin{center}
    \mygraphics{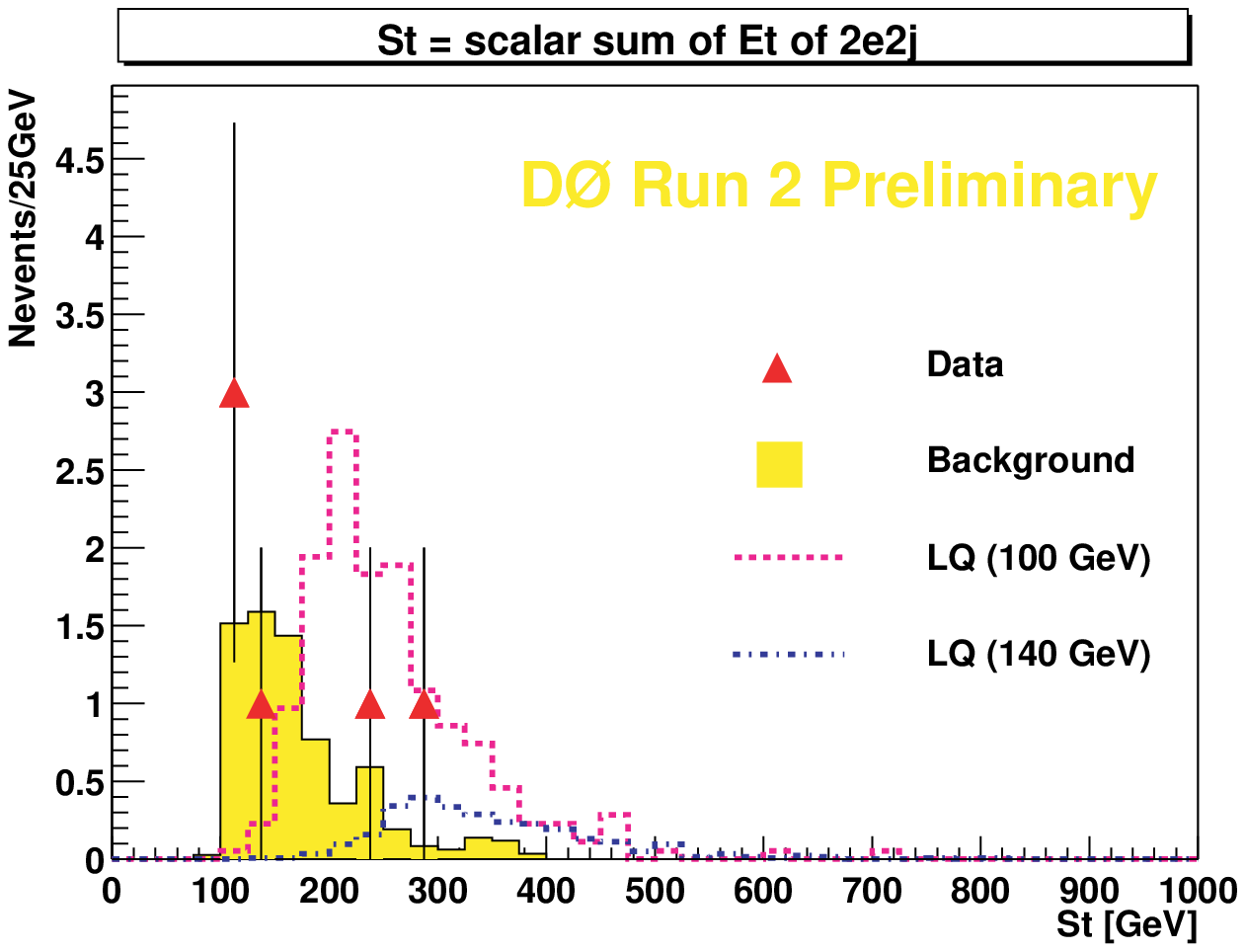}
    \caption{Distribution of the scalar sum of \Et\ seen in the
      \Dzero\ detector for $eejj$ events.}
    \label{fig:d0lq}
  \end{center}
\end{minipage}
\end{figure}

First attempts at searching for new physics signals have already been
made by the \Dzero\ collaboration\cite{bsmichep}.
Gauge mediated models of supersymmetry 
can give scenarios where the lightest
supersymmetric particle is the gravitino (the partner of the graviton)
and the next to lightest supersymmetric particle is a neutralino,
$\chi_1^0$ (one of the partners of the neutral gauge bosons and
Higgses) or a slepton (one of the lepton partners). 
These lead to decay modes of the type:
\begin{displaymath}
  \ppb \ra \mathrm{gauginos} 
  \ra \Wboson /\Zboson /\gamma + \chi_1^0 \chi_1^0
  \ra \gamma \gamma + \tilde{G} \tilde{G} + X.
\end{displaymath}
The gravitinos, $\tilde{G}$, do not interact in the detector and are
detected as missing energy.
Results of the search are shown in Figure \ref{fig:d0gmsusy}.
Although the sensitivity with the $\sim$10 \invpb\ collected so far is
too small to exclude any of the SUSY parameter space,
an approximately model independent lower limit for the cross-section
of this process has been set at 0.9 pb.

Another possibility for extensions of the SM is lepto-quark models,
where new particles exist that carry both quark and lepton quantum
numbers. \Dzero\ has searched for such particles in the 2-electron +
2-jet channel, unfortunately with no success (see Figure
\ref{fig:d0lq}). This allows a limit on the presumed lepto-quark mass
of, $M_{LQ}$ $>$ 113 \GeV\ to be set at the 95\% CL (assuming the
branching ratio to this mode is 1).

\begin{figure}[ht]
\begin{center}
  \mygraphics{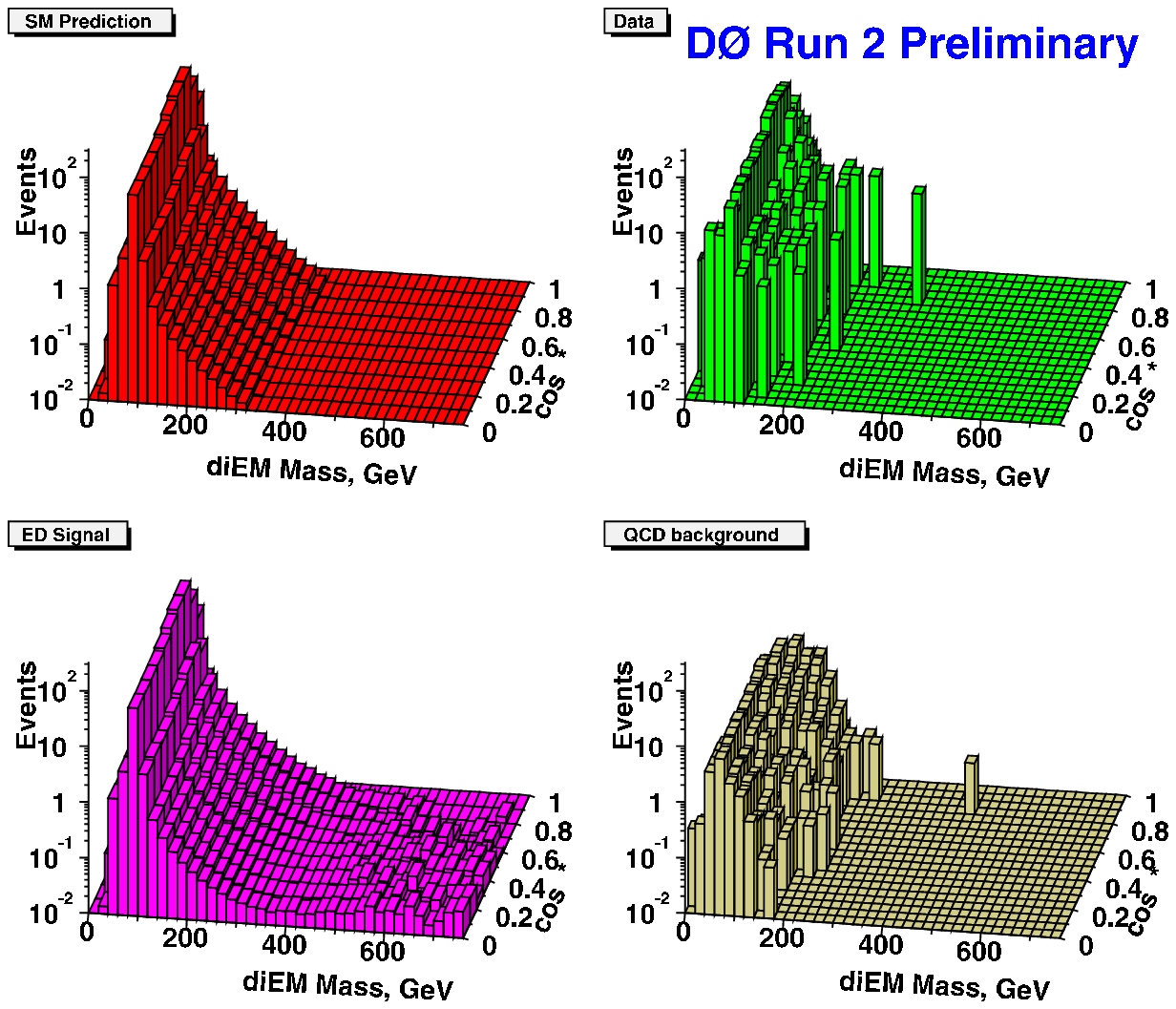}
  \caption{Distributions of diEM invariant mass vs. center-of-mass
    scattering angle in the \Dzero\ large extra dimensions search.
    The SM prediction is shown in the upper-left, 
    the \Dzero\ data in the upper-right,
    the prediction for an extra dimensions signal in the lower-left 
    and the QCD background to the search in the lower-right.}
  \label{fig:d0led}
\end{center}
\end{figure}

Finally, \Dzero\ has also looked for gravitons arising in theories
with large extra dimensions. Such particles can decay to to \epem\ or
$\gamma\gamma$ and interfere with the SM production mechanisms for
these final states. A two-dimensional distribution of diEM (electrons
and photons are not distinguished) invariant mass vs. the cosine of
the scattering angle in the center-of-mass frame of the hard scatter
is shown in Figure \ref{fig:d0led}.
Again, the data distribution is indistinguishable from the SM
prediction 
allowing a limit on the fundamental scale of gravity to be set at
\MS\ $>$ 0.82 \TeV , in Hewett's convention\cite{hewett}.
This compares favorably with the limit set in the \Dzero\ Run I search
-- \MS\ $>$ 1.2 \TeV\ (Ref. \citenum{runiled}).

\section{Conclusions}
From QCD studies, to electro-weak precision measurements, to probes of
CP violation and searches for as-yet undiscovered particles, the Run
II physics menu is full of interesting topics.
We have seen that world-best levels sensitivity are expected for a wide
range of important measurements in Run II.
First physics results presented at the ICHEP
conference in July, 2002 indicate that we are well on the way to
achieving these goals,
with detector performances beginning to approach design
specifications and sophisticated analysis techniques being tuned up.
Of course, with such an exciting program, things are not standing
still at Fermilab. Accelerator and detector performances are
continuing to improve.
In fact, with the steady running we have enjoyed through the fall we
should have a data 
sample of 50 \invpb\ or more by the end of the year and conceivably
250 \invpb\ (more than twice the Run I sample) in time for the summer.
Further in the future, a new series of upgrades to the machine and the
detectors should allow us to collect 15 \invfb\ by the time LHC starts
up. With data sets of this size, a low mass Higgs is well within our
grasp. 
The best is clearly yet to come!

\section{Acknowledgments}
I would like to acknowledge all of the CDF and \Dzero\ speakers at ICHEP
from whom I stole material shamelessly. I also profited from the good
advice of B. Ashmanskas, G. Blazey, G. Brooijmans, J. Dittmann, Y. Gershtein,
A. Goshaw, J. Houston, J. Krane, G. Landsberg, N. Lockyer, M. Narain,
G. Steinbr\"{u}ck, J. Womersley and S. Worm. Finally, I am happy to
thank the SLAC Summer Institute organizers for providing such a
stimulating conference, and for putting up with all my last-minute
requests.



\begin{thebibliography}{10}
\bibitem{alice}
	``What I tell you three times is true'' \\
	from {\it The Hunting of the Snark}, by Lewis Carrol.
\bibitem{smhewett}
	For a review see,
	J. Hewett, {\it The Standard Model and Why We Believe It},
	\hepref{ph}{9810316}.
\bibitem{bernreuther}
	For a review see,
	W. Bernreuther, {\it CP Violation and Baryogenesis},
	\hepref{ph}{0205279}.
\bibitem{topdisc}
	F.Abe, \etal\ (CDF),
	\Journal{\PRL}{74}{2626}{1995}. \\
	S.Abachi, \etal\ (\Dzero ),
	\Journal{\PRL}{74}{2632}{1995}.
\bibitem{wuerthwein}
	F. W\"{u}rthwein,
	  {\it B Physics Studies at the Tevatron},
	  these proceedings.
\bibitem{tevupgrade}
	See the {\it Run II Handbook}, available at \\
	\WWWAddr{http://www-bd.fnal.gov/runII/index.html} \\
	for a more detailed description.
\bibitem{ichep02}
	{\it International Conference on High Energy Physics},
	July 24-31, 2002,
	Amsterdam, the Netherlands,
	\WWWAddr{http://www.ichep02.nl}.
\bibitem{cdfweb}
	Information about the CDF detector, collaboration and
	physics results can be found in the CDF WWW pages:
	\WWWAddr{http://www-cdf.fnal.gov/}.
\bibitem{d0web}
	Information about the \Dzero\ detector, collaboration and
	physics results can be found in the \Dzero\ WWW pages:
	\WWWAddr{http://www-d0.fnal.gov/}.
\bibitem{bedeschi}
	F. Bedeschi, {\it First CDF Run II Results}, \\
	talk presented at the
	{\it International Conference on High Energy Physics},
	July 24-31, 2002,
	Amsterdam, the Netherlands,
\bibitem{narain}
	M. Narain, {\it Results from the \Dzero\ Experiment at the
	Tevatron}, \\
	talk presented at the
	{\it International Conference on High Energy Physics},
	July 24-31, 2002,
	Amsterdam, the Netherlands,
\bibitem{runiiWG}
	Information about all the Tevatron Run II Working Groups can
	  be found off of the Fermilab theory group web page:
	  \WWWAddr{http://theory.fnal.gov/}.
\bibitem{snowmass}
	See the Snowmass web page:
	\WWWAddr{http://snowmassserver.snowmass2001.org}.
\bibitem{tevbwg}
	K. Anikeev, \etal ,
	  {\it B Physics at the Tevatron: Run II and Beyond},
	  \hepref{ph}{0201071}.
\bibitem{dittman}
	J. Dittman, {\it Photon and Jet Physics at CDF}, \\
	M. Zielinski, {\it Jet and Photon Physics at \Dzero }, \\
	talks presented at the
	{\it International Conference on High Energy Physics},
	July 24-31, 2002,
	Amsterdam, the Netherlands,
\bibitem{glenzinski}
	D. Glenzinski, {\it Electroweak Prospects for Tevatron Run II},\\
	talk presented at the
	{\it International Conference on High Energy Physics},
	July 24-31, 2002,
	Amsterdam, the Netherlands,
\bibitem{iashvilli}
	I. Iashvilli, {\it Top Quark Physics at the Tevatron}, \\
	talk presented at the
	{\it International Conference on High Energy Physics},
	July 24-31, 2002,
	Amsterdam, the Netherlands,
\bibitem{lhcew}
	S. Haywood, \etal ,
	{\it Electroweak Physics},
	\hepref{ph}{0003275}.
\bibitem{lhctop}
	M. Beneke, \etal ,
	{\it Top Quark Physics},
	\hepref{ph}{0003033}.
\bibitem{lepewwg}
	LEP Electroweak Working Group fits, 
	M. Gr\"{u}newald, {\it Electroweak Physics},
	\hepref{ex}{0210003}. \\
	talk presented at the
	{\it International Conference on High Energy Physics},
	July 24-31, 2002,
	Amsterdam, the Netherlands,
\bibitem{tevewwg}
	U. Baur, R.K. Ellis and D. Zeppenfeld,
	  {\it QCD and Weak Boson Physics at Run II}
	  \FermiPub{00}{297}.
\bibitem{pdg}
	K. Hagiwara, \etal\ (Particle Data Group),
	\Journal{\PhysRev}{D66}{}{2002}.
\bibitem{snowew}
	U. Baur, \etal ,
	{\it Present and Future Electroweak Precision Measurements and
	the Indirect Determination of the Mass of the Higgs Boson},
	\hepref{ph}{0202001}.
\bibitem{cdftpol}
	T. Affolder, \etal\ (CDF),
	\Journal{\PRL}{84}{216}{2000}.
\bibitem{cdf1top}
	D. Acosta, \etal\ (CDF),
	\Journal{\PhysRev}{D65}{091120}{2002}.
\bibitem{lephiggs}
	The LEP Working Group for Higgs Boson Searches,
	{\it Search for the Standard Model Higgs Boson at LEP},
	LHWG Note/2002-01. \\
	(\WWWAddr{http://lephiggs.web.cern.ch/LEPHIGGS/www/Welcome.html})
\bibitem{tevhiggswg}
	M. Carena, J.S. Conway, H.E Haber and J.D. Hobbs,
	  {\it Report of the Tevatron Higgs Working Group},
	  \hepref{ph}{0010338}.
\bibitem{tevsugrawg}
	V. Barger, C.E.M. Wagner, T. Kamon, E. Flattum, T. Falk and
	  X. Tata,
	  {\it Report of the SUGRA Working Group for Run II at the
	  Tevatron},
	  \hepref{ph}{0003154}.
\bibitem{tevbmssmwg}
	S. Ambrosanio, \etal ,
	  {\it Report of the Beyond the MSSM Subgroup for the Tevatron
	  Run II SUSY/Higgs Workshop},
	  \hepref{ph}{0006162}.
\bibitem{tevrpvwg}
	B. Allanach, \etal ,
	  {\it Searching for R-Parity Violation at Run II of the Tevatron}
	  \hepref{ph}{9906224}.
\bibitem{tevgmsusywg}
	R. Culbertson, S.P. Martin, J. Qiang and S. Thomas,
	  {\it Low-Scale and Gauge-Mediated Supersymmetry Breaking
	  at the Fermilab Tevatron Run II},
	  \hepref{ph}{0008070}.
\bibitem{susy}
	J. Wess and J. Bagger,
	\Book{Introduction to Supersymmetry, 2nd ed.}{}%
             {Princeton University Press}{1992}; \\
	H.P. Nilles, \Journal{\PhysRep}{110}{1}{1984}; \\
	H. Haber and G. Kane, \Journal{\PhysRep}{117}{75}{1985}.
\bibitem{techni}
	For a review see, \eg , R.S. Chivukula, R. Rosenfeld,
	E.H. Simmons and J. Terning in, 
	\Book{Electroweak Symmetry Breaking and New Physics at the TeV %
	      Scale}{edited by T.L. Barlow, S. Dawson, H.E. Haber and %
	      J.L. Siegrist}%
             {World Scientific}{1996}; \\
	E. Eichten, K, Lane and J. Womersley, 
	\Journal{\PhysLett}{B405}{305}{1997}; \\
	K. Lane, \Journal{\PhysRev}{D60}{075007}{1999}.
\bibitem{ledorig}
	N. Arkani-Hamed, S. Dimopoulos, G. Dvali,
	\Journal{\PhysLett}{B429}{263}{1998}.
\bibitem{raretop}
	G. Eilam, J. Hewett and A. Soni,
	\Journal{\PhysRev}{D44}{1473}{1991}. \\
	Erratum, \Journal{\PhysRev}{D59}{039901}{1999}.
\bibitem{zprime}
	for a recent review see,
	A. Leike, \Journal{\PhysRep}{317}{143}{1999}.
\bibitem{lepsusy}
	The LEP Higgs Working Group,
	{\it Searches for the Neutral Higgs Bosons of the MSSM:
	  Preliminary Combined Results Using LEP Data Collected 
	  at Energies up to 209 GeV},
	\hepref{ex}{0107030}.
\bibitem{atlastdr}
	{\it ATLAS Detector and Physics Performance Technical Design
	Report},
	LHCC 99-14/15 (1999). \\
	(\WWWAddr{http://atlas.web.cern.ch/Atlas/GROUPS/PHYSICS/TDR/}\\
	\WWWAddr{access.html})
\bibitem{orejudos}
	W. Orejudos, {\it Minireview on Other Signatures of New
	  Physics at the Tevatron}, \\
	talk presented at the
	{\it International Conference on High Energy Physics},
	July 24-31, 2002,
	Amsterdam, the Netherlands.
\bibitem{ledpred}
	K. Cheung and G. Landsberg,
	\Journal{\PhysRev}{D62}{076003}{2000}
	(\hepref{ph}{9909218}).
\bibitem{tovey}
	D. Tovey, {\it Searches for New Physics at the LHC}, \\
	talk presented at the
	{\it International Conference on High Energy Physics},
	July 24-31, 2002,
	Amsterdam, the Netherlands.
\bibitem{cdfraret}
	F. Abe, \etal\ (CDF),
	\Journal{\PRL}{80}{2525}{1998}.
\bibitem{lcexotic}
	T. Abe, \etal ,
	{\it Linear Collider Physics Resource Book for Snowmass 2001,
	  Part 3: Studies of Exotic and Standard Model Physics},
	\hepref{ex}{0106057}.\\
	(\WWWAddr{http://www.slac.stanford.edu/grp/th/LCBook/})
\bibitem{bsmichep}
	G. Bernardi, {\it Minireview on Low-Scale Gravity and Extra
	  Dimensions at HERA, LEP and the Tevatron},\\
	A. Connolly, {\it Search for MSSM Higgses at the Tevatron}, \\
	V. Zutshi, {\it Search for SUSY at the Tevatron}, \\
	talks presented at the
	{\it International Conference on High Energy Physics},
	July 24-31, 2002,
	Amsterdam, the Netherlands.
\bibitem{hewett}
	J. Hewett, \Journal{\PRL}{82}{4765}{1999}.
\bibitem{runiled}
	B. Abbott, \etal\ (\Dzero ),
	\Journal{\PRL}{86}{1156}{2001}.

\end{thebibliography}
\end{document}